%% file: hatp17_astroph.tex
\documentclass[apjl]{emulateapj}
\usepackage{ifthen}

\input{newcommand_gen.tex}

\input{newcommand_spec.tex}

\newcommand{\hatcurisoshort}{YY}
\newcommand{\hatcurisofull}{Yonsei-Yale (YY)}
\newcommand{\hatcurisocite}{yi:2001}
\newcommand{\hatcurlumind}{\arstar}
\newcommand{\hatcurjhkfilset}{ESO}
\newboolean{emulateapj}
\setboolean{emulateapj}{true}

\newboolean{rvtablelong}
\setboolean{rvtablelong}{true}

\newboolean{astroph}
\setboolean{astroph}{true}
\shortauthors{Howard et al.}
\shorttitle{\hatcur\lowercase{b}}
\ifthenelse{\boolean{emulateapj}}{
    \newcommand{\titledag}{$\dagger$}
}{
    \newcommand{\titledag}{\dagger}
}

\begin{document}

%% Titlepage
\title{\hatcurbc: A Transiting, Eccentric, Hot Saturn and a Long-period, Cold Jupiter  \altaffilmark{\titledag}}

%%%  %%%
%% Authors
\author{ 
    A.~W.~Howard\altaffilmark{1,2},
    G.~\'A.~Bakos\altaffilmark{3,4},
    J.~Hartman\altaffilmark{3},
    G.~Torres\altaffilmark{3},
    A.~Shporer\altaffilmark{5},
    T.~Mazeh\altaffilmark{6},
    G\'eza~Kov\'acs\altaffilmark{7},
    D.~W.~Latham\altaffilmark{3},
    R.~W.~Noyes\altaffilmark{3},
    D.~A.~Fischer\altaffilmark{8},
    J.~A.~Johnson\altaffilmark{9},
    G.~W.~Marcy\altaffilmark{1},
    G.~A.~Esquerdo\altaffilmark{3},
    B.~B\'eky\altaffilmark{3},
    R.~P.~Butler\altaffilmark{10},
    D.~D.~Sasselov\altaffilmark{3},
    R.~P.~Stefanik\altaffilmark{3},
    G.~Perumpilly\altaffilmark{3},
    J.~L\'az\'ar\altaffilmark{11},
    I.~Papp\altaffilmark{11},
    P.~S\'ari\altaffilmark{11}
}

\altaffiltext{1}{Department of Astronomy, University of California,
    Berkeley, CA}

\altaffiltext{2}{Townes Postdoctoral Fellow; 
         howard@astro.berkeley.edu}

\altaffiltext{3}{Harvard-Smithsonian Center for Astrophysics,
    Cambridge, MA; email: gbakos@cfa.harvard.edu}

\altaffiltext{4}{NSF Fellow}

\altaffiltext{5}{
	LCOGT, 6740 Cortona Drive, Santa Barbara, CA, \& Department of Physics,
	Broida Hall, UC Santa Barbara, CA}

\altaffiltext{6}{
	School of Physics and Astronomy, Raymond and Beverly Sackler Faculty of
	Exact Sciences, Tel Aviv University, Tel Aviv 69978, Israel}

\altaffiltext{7}{Konkoly Observatory, Budapest, Hungary}

\altaffiltext{8}{Astronomy Department, Yale University,
	New Haven, CT}

\altaffiltext{9}{California Institute of Technology, Department of
    Astrophysics, MC 249-17, Pasadena, CA}

\altaffiltext{10}{Department of Terrestrial Magnetism, Carnegie
	Institute of Washington, DC}

\altaffiltext{11}{Hungarian Astronomical Association, Budapest, 
    Hungary}

\altaffiltext{$\dagger$}{%%
    Based in part on observations obtained at the W.~M.~Keck
    Observatory, which is operated by the University of California and
    the California Institute of Technology. Keck time has been
    granted by NOAO and NASA.
}

%% EOF authors

% #####################################################################
%% abstract
\begin{abstract}
\setcounter{footnote}{11} 
We report the discovery of \hatcurbc{}, a multi-planet system with an 
inner transiting planet in a short-period, eccentric orbit 
and an outer planet in a 4.8\,yr, nearly circular orbit.  
The inner planet, \hatcurb, transits the bright
V = \hatcurCCtassmv\ \hatcurISOspec\ dwarf star \hatcurCCgsc, 
with an orbital period
$P=\hatcurLCP$\,d, orbital eccentricity $e = \hatcurRVeccen$,  
transit epoch $T_c = \hatcurLCT$
(BJD\footnote{Barycentric Julian dates throughout the paper are
calculated from Coordinated Universal Time (UTC)}), and transit
duration \hatcurLCdur\,d. 
\hatcurb\ has a mass of
\hatcurPPmlong\,\mjup\ and radius of \hatcurPPrlong\,\rjup\ yielding a
mean density of \hatcurPPrho\,\gcmc.
This planet has a relatively low equilibrium temperature 
in the range 780--927\,K, making it an attractive target for 
follow-up spectroscopic studies.  
The outer planet, \hatcurc, has a significantly longer orbital period 
$P_2 = \hatcurcLCP$\,d and a 
minimum mass $m_2\sin i_2 = \hatcurcPPmlong\,\mjup$.  
The orbital inclination of \hatcurc\ is unknown as 
transits have not been observed and may not be present. 
The host star has a mass of   
\hatcurISOm\,\msun, radius of \hatcurISOr\,\rsun, effective
temperature \hatcurSMEteff\,K, and metallicity $\feh = \hatcurSMEzfeh$. 
\hatcur\ is the second multi-planet system detected from ground-based 
transit surveys.  
\setcounter{footnote}{0}
\end{abstract}

% #####################################################################
%% keywords
\keywords{
	planetary systems ---
	stars: individual (\hatcur{}, \hatcurCCgsc{}) 
	techniques: spectroscopic, photometric
}

%% EOF keywords
%% EOF titlepage

% #####################################################################
%% Introduction
\section{Introduction}
\label{sec:introduction}
%% EOF introduction

With nearly 100 confirmed transiting extrasolar planets (TEPs) known,
many studies of planetary properties now focus 
on the statistical distributions of and correlations between 
planetary parameters.
Individual TEPs still remain extraordinarily valuable, 
particularly if they have properties that 
exemplify an important subgroup of planets 
and orbit stars that are bright 
enough for meaningful follow-up observations.  
Such iconic well-studied planets include
HD~209458b \citep{dc:2000,henry:2000}, 
HD~189733 \citep{bouchy:2005}, 
GJ~436b \citep{gillon:2007,butler:2004}, 
HAT-P-13b,c \citep{bakos:2010},
WASP-12b \citep{hebb:2009}, 
and GJ~1214b \citep{dc:2009}.  
The \hatcur\ system has at least two unusual properties 
compared to the ensemble of known TEPs 
and may serve as an exemplar for planets with these 
properties.   The atmosphere of 
\hatcurb\ is relatively cool for a TEP and 
\hatcurc\ is one of only two long-period planets found to orbit 
a TEP host.  

The distribution of TEPs discovered by ground-based transit 
surveys is biased toward 
large planets orbiting faint early type stars in short period orbits.  
Each of these biases stems from the observational selection effects 
of the surveys that have detected the majority of TEPs: 
the deep transits of large planets are easier to detect; 
early type stars dominate magnitude limited surveys; 
and short-period orbits have higher \textit{a prioiri} 
transit probabilities and significantly larger probabilities of 
detection in a ground-based survey limited to one or two 
observing sites.  
The overabundance of detected short period planets has skewed 
our perception of the atmospheric properties of extrasolar jovian planets.  
Planets in $P \sim 3$\,d orbits ($a \sim 0.04$) experience intense 
interactions with the radiation and tides
of their host stars.  
The atmospheres of many of these planets are inflated beyond  
the radii predicted by theoretical models  \citep{fortney:2007}.  
To understand cooler planets, 
which we know from radial velocity (RV) surveys represent 
the vast majority of gas giants \citep{wright:2009}, 
we must study planets orbiting progressively 
further from their host stars.   

With such a scarcity of cool planets, 
\hatcurb\ is a valuable probe of the planetary mass--radius relationship 
and additional properties through follow-up observations.  
Together, the relatively long orbital period and later spectral type of the host star 
yield an incident stellar flux that 
is 1--2 orders of magnitude lower than the flux received by most 
detected TEPs \citep{kovacs:2010}. 
The host star is also relatively bright, $V = \hatcurCCtassmv$, 
making follow-up atmospheric studies conceivable.
While the \textit{Kepler} mission \citep{borucki:2010}
has been extraordinarily successful 
in the detection of TEPs, the vast majority of its discovered planets 
orbit faint stars; only 1.5\% of the $\sim10^5$ stars being surveyed  
\citep{batalha:2010} are brighter than 11th magnitude 
(Kepler magnitude).  
We predict that \hatcurb\ will be among the small number of 
well-studied cooler ($\teff < 1000$\,K) TEPs.

Prior to this announcement, only one TEP discovered by a 
ground-based photometric survey is in a confirmed multi-planet system.  
HAT-P-13 \citep{bakos:2010}
has a hot Jupiter inner planet and highly eccentric super-Jupiter 
outer planet with an orbital period of ~450~d.  
The outer planet, HAT-P-13c, was detected only in RV measurements  
and has not been shown to transit.  
The system reported here, \hatcur, is now the second TEP 
discovered by a ground-based 
transit survey with a confirmed second planet. 
The relatively low rate of detected planet multiplicity among TEPs
discovered from by ground may be skewed by the lack of
long-term RV and photometric monitoring for most TEP host stars.
Measuring the rate of planet multiplicity among 
stars hosting a hot Jupiter will probe 
the dynamical histories and migrations mechanisms of 
hot Jupiters \citep[see, e.g.,][]{wu:2007}.
Multi-planet systems are significantly more common among 
RV-detected systems; \cite{wright:2009} find that 28\% of known 
planet host stars are multi-planet systems.  

Several multi-planet systems have also been discovered from space.  
Corot-7 is thought to host two short-period 
super-Earths \citep{leger:2009,queloz:2009}, one of which transits.
The \textit{Kepler} mission recently announced five candidate 
systems with multiple transiting planets and is poised to announce 
additional systems \citep{steffen:2010}.

The Hungarian-made Automated Telescope Network
\citep[HATNet;][]{bakos:2004} survey has been one of the main
contributors to the discovery of TEPs.  In operation since 2003, it
has now covered approximately 14\% of the sky, searching for TEPs
around bright stars ($8\lesssim I \lesssim 14.0$).  HATNet operates
six wide-field instruments: four at the Fred Lawrence Whipple
Observatory (FLWO) in Arizona, and two on the roof of the hangar
servicing the Smithsonian Astrophysical Observatory's Submillimeter
Array, in Hawaii.  Since 2006, HATNet has announced and published 16
TEPs.  In this work we report our seventeenth discovery, around the
relatively bright star previously known as \hatcurCCgsc{}.

The layout of the paper is as follows. In \refsecl{obs} we report the
detection of the photometric signal and the follow-up spectroscopic
and photometric observations of \hatcur{}.  In \refsecl{analysis} we
describe the analysis of the data, beginning with the determination of
the stellar parameters, continuing with a discussion of the methods
used to rule out nonplanetary, false positive scenarios which could
mimic the photometric and spectroscopic observations, and finishing
with a description of our global modelling of the photometry and radial
velocities.  In \refsecl{discussion} we discuss implications of this 
discovery, compare our results with recent theoretical models of TEPs, 
and consider possible follow-on observations.

% #####################################################################
\section{Observations}
\label{sec:obs}

% =====================================================================
%% Photometric detection
\subsection{Photometric detection}
\label{sec:detection}

The transits of \hatcurb{} were detected with the HAT-5 telescope in
Arizona, the HAT-8 telescope in Hawaii, and with the Wise-HAT (WHAT)
telescope at Wise Observatory in Israel \citep{shporer:2009}. The
region around \hatcurCCgsc{}, a field internally labeled as
\hatcurfield, was observed on a nightly basis between 2005 May 8 and
2005 October 24, whenever weather conditions permitted. We gathered
9686 exposures of 5 minutes duration at a 5.5 minute cadence. 
Each image
contained approximately 85,000 stars down to $I\sim14.0$. For the
brightest stars in the field, we achieved a per-image photometric
precision of 5\,mmag. The star is also located in the overlapping
field \hatcurfieldtwo, which was observed with the HAT-6 telescope in
Arizona and the WHAT telescope in Israel between 2004 June 4 and
2004 November 10, and between 2005 July 3 and 2005 July 15. Altogether
4882 exposures of 5 minutes duration at 5.5 minute cadence 
were gathered for this field.

%% ----------------
\begin{figure}[!ht]
\plotone{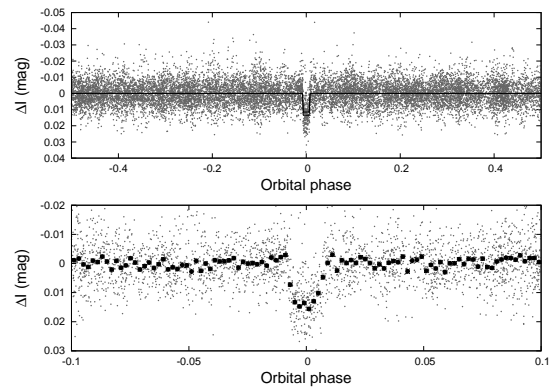}
\caption{
	Unbinned (top) and binned (bottom) \lc{}s of \hatcur{} 
	including all 14,000 instrumental
        \band{I} 5.5 minute cadence measurements obtained with the
        HAT-5, HAT-6, and HAT-8 telescopes of HATNet and with the
        WHAT telescope (see the text for details), and folded with
        the period $P = \hatcurLCPprec$\,days (resulting from the
        global fit described in \refsecl{analysis}). The solid line
        shows the ``P1P3'' transit model fit to the light curve
        (\refsecl{globmod}).
\label{fig:hatnet}}
\end{figure}
%% ----------------

The calibration of the HATNet and WHAT frames was carried out using
standard photometric procedures.  The calibrated images were then
subjected to star detection and astrometry, as described in
\cite{pal:2006}.  Aperture photometry was performed on each image at
the stellar centroids derived from the Two Micron All Sky Survey
\citep[2MASS;][]{skrutskie:2006} catalog and the individual
astrometric solutions.  The resulting \lcs\ were decorrelated (cleaned
of trends) using the External Parameter Decorrelation \citep[EPD;
  see][]{bakos:2009} technique in ``constant'' mode and the Trend
Filtering Algorithm \citep[TFA; see][]{kovacs:2005}.  The \lcs{} were
searched for periodic box-shaped signals using the Box Least-Squares
\citep[BLS; see][]{kovacs:2002} method.  We detected a significant
signal in the \lc{} of \hatcurCCgsc{} (also known as
\hatcurCCtwomass{} and TYC 2717-417-1; 
$\alpha = \hatcurCCra$, $\delta = \hatcurCCdec$;
J2000; V=\hatcurCCtassmv{}; \citealp{droege:2006}), with an apparent
depth of $\sim\hatcurLCdip$\,mmag, and a period of
$P=\hatcurLCPshort$\,days (see \reffigl{hatnet}).  The drop in
brightness had a first-to-last-contact duration, relative to the total
period, of $q = \hatcurLCq$, corresponding to a total duration of $Pq =
\hatcurLCdurhr$~hr (see \reffigl{hatnet}). We note that the transits
were only detected from the observations of field \hatcurfield, and
were not detected in the observations of field \hatcurfieldtwo{}.

% =====================================================================
\subsection{Reconnaissance Spectroscopy}
\label{sec:recspec}

As is routine in the HATNet project, all candidates are subjected to
careful scrutiny before investing valuable time on large
telescopes. This includes spectroscopic observations at relatively
modest facilities to establish whether the transit-like feature in the
light curve of a candidate might be due to astrophysical phenomena
other than a planet transiting a star. Many of these false positives
are associated with large radial-velocity variations in the star (tens
of \kms) that are easily recognized.

% TODO: Add discussion of TRES and/or Du Pont if used, no published planet
%       has made use of either facility to date, so it will need to be 
%       generated from scratch.

One of the tools we have used for this purpose is the
Harvard-Smithsonian Center for Astrophysics (CfA) Digital Speedometer
\citep[DS;][]{latham:1992}, an echelle spectrograph mounted on the
\flwos\ telescope. This instrument delivers high-resolution spectra
($\lambda/\Delta\lambda \approx 35,\!000$) over a single order
centered on the \ion{Mg}{1}\,b triplet ($\sim$5187\,\AA), with
typically low signal-to-noise (S/N) ratios that are nevertheless
sufficient to derive radial velocities (RVs) with moderate precisions
of 0.5--1.0\,\kms\ for slowly rotating stars.  The same spectra can be
used to estimate the effective temperature, surface gravity, and
projected rotational velocity of the host star, as described by
\cite{torres:2002b}.  With this facility we are able to reject many
types of false positives, such as F dwarfs orbited by M dwarfs,
grazing eclipsing binaries, or triple or quadruple star
systems. Additional tests are performed with other spectroscopic
observations described in the next section.

For \hatcur{} we obtained eight observations with the DS between
September and November of 2007.  The velocity measurements showed an
rms residual of \hatcurDSrvrms\,\kms, consistent with no detectable RV
variation within the precision of the measurements. All spectra were
single-lined, i.e., there is no evidence for additional stars in the
system.  The atmospheric parameters we infer from these observations
are the following: effective temperature $\teffstar =
\hatcurDSteff\,K$, surface gravity $\loggstar = \hatcurDSlogg$ (log
cgs), and projected rotational velocity $\vsini = \hatcurDSvsini\,\kms$. 
The effective temperature corresponds to an \hatcurISOspec\ dwarf star. 
The mean heliocentric RV of \hatcur\ is
$\gamma_{\rm RV} = \hatcurDSgamma$\,\kms.

% =====================================================================
\subsection{High resolution, high S/N spectroscopy}
\label{sec:hispec}

Given the significant transit detection by HATNet, and the encouraging
DS results that rule out obvious false positives, we proceeded with
the follow-up of this candidate by obtaining high-resolution, high-S/N
spectra to characterize the RV variations, and to refine the
determination of the stellar parameters. For this we used HIRES
\citep{vogt:1994} on the Keck~I telescope located on Mauna
Kea, Hawaii, between 2007 October and 2010 April.  The width of the
spectrometer slit was $0\farcs86$, resulting in a resolving power of
$\lambda/\Delta\lambda \approx 55,\!000$, with a wavelength coverage
of $\sim$3800--8000\,\AA\@.

\begin{figure*}% [h]
\plotone{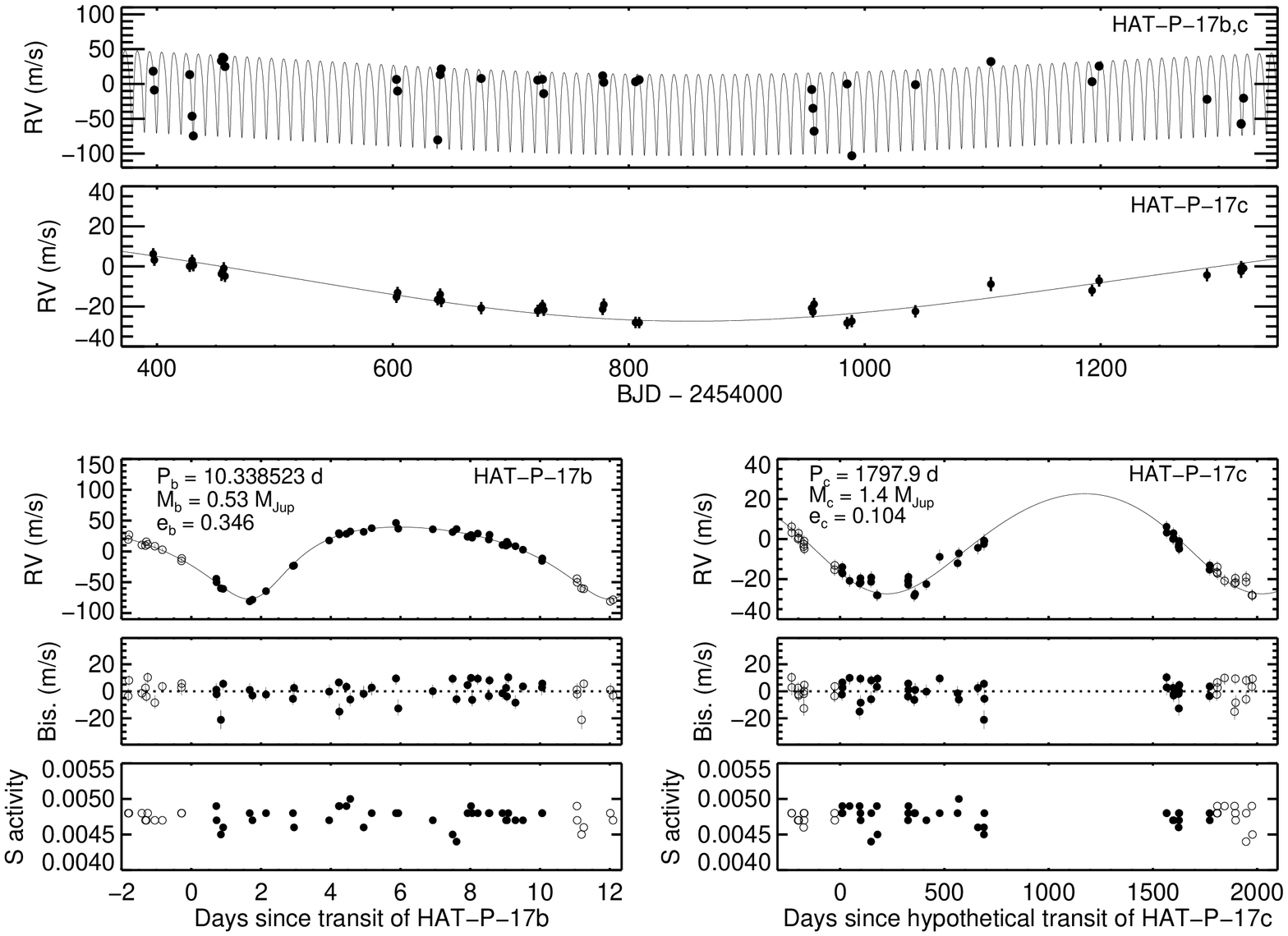}
\caption{
        {\em Top:}  Keck/HIRES RV measurements for
        \hbox{\hatcur{}} shown as a function of BJD, along with our
        best-fit 2-planet model (see \reftabl{planetparam}). The
        center-of-mass velocity has been subtracted.
        {\em Second from top:} Same as top panel except the 
        RV model of the inner planet has been subtracted 
        from the data and the model, revealing the orbit of the outer 
        planet.  The rms variation of the
        residuals to the two-planet model is 3.07\,\ms, 
        requiring a jitter of
        $\hatcurRVjitter$\,\ms\ added in quadrature to the
        individual errors to yield a reduced $\chisq$ of 1.0. The
        error-bars in this panel have been inflated accordingly.
        {\em Third row:} RV measurements phased to the 
        orbital periods of the inner planet (left) and 
        the outer planet (right).  
        In each plot the orbit of the other planet has been 
        removed.  
        {\em Fourth row:} Bisector spans (BS), with the mean value
        subtracted, phased at the period of the inner planet (left) 
        and the outer planet (right). The
        measurement from the template spectrum is included (see
        \refsecl{bisec}).
        {\em Fifth row:} Relative chromospheric activity index $S$
        measured from the Keck spectra, phased at the period of the
        inner planet (left) and the outer planet (right).
	Note the different vertical scales of the panels. Observations
        shown twice are represented with open symbols.
\label{fig:rvbis}}
\end{figure*}

We obtained 32 HIRES exposures with an iodine cell mounted directly 
in front of the spectrometer entrance slit.  
The dense set of molecular absorption lines imprinted 
on the stellar spectra provide a robust wavelength fiducial 
against which Doppler shifts are measured, 
as well as strong constraints on the shape of the spectrometer instrumental 
profile at the time of each observation \citep{marcy:1992,valenti:1995}.
An additional exposure was taken
without the iodine cell, for use as a template in the reductions.
Relative RVs in the solar system barycentric frame were derived as
described by \cite{butler:1996}, incorporating full modeling of the
spatial and temporal variations of the instrumental profile.  
These measurements have typical uncertainties of 1.5--2.0~\ms\ 
for spectra with per-pixel signal-to-noise ratios of 100--150.  
HIRES measurements of late G and early K dwarf stars have achieved 
long term stability of 1.5--2.0~\ms\ on standard stars, 
including noise from systematic and astrophysical sources 
\citep{howard:2010}.
The RV measurements and their uncertainties are listed in \reftabl{rvs}. 
The period-folded data, along with our best fit described below in
\refsecl{analysis}, are displayed in \reffigl{rvbis}.

%% --------------------------------------------------------------------
%% --------------------------------------------------------------------

In the same figure we also show the relative $S$ index, which is a
measure of the chromospheric activity of the star derived from the
flux in the cores of the \ion{Ca}{2} H and K lines.  This index was
computed following the prescription given by \citet{vaughan:1978},
after matching each spectrum to a reference spectrum using a
transformation that includes a wavelength shift and a flux scaling
that is a polynomial as a function of wavelength. The transformation
was determined on regions of the spectra that are not used in
Note that our relative $S$ index has not been calibrated to the scale
of \citet{vaughan:1978}. We do not detect any significant variation of
the index correlated with the orbital phase of either planet; 
such a correlation might have
indicated that the RV variations could be due to stellar activity,
casting doubt on the planetary nature of the candidates. 

In addition, we computed an $S_{\rm HK}$ index calibrated 
to the Mt.\ Wilson scale, permitting comparisons with 
calibrated activity measurements of other stars \citep{knutson:2010}. 
We find $S_{\rm HK} = 0.162$ (median of all HIRES measurements) and 
\logrhk\ = $-$5.039 (median).
These measurements employ the techniques described in 
\cite{isaacson:2010}, calibrated on 1500 stars observed by the 
California Planet Survey (CPS).  
We used $B-V = 0.83$ estimated from \teff\ using the linear transformation 
between those quantities in \cite{valenti:2005}.

%% --------------------------------------------------------------------
%% Note: there are two possible versions for this table: one with
%% BJD, RB, RVerr, BS, BSerr, and another one with the former columns
%% plus S and Serr. For the first form it is OK to use the deluxetable
%% environment. 
%% With the second form we need deluxetable*.
%%
\ifthenelse{\boolean{emulateapj}}{
    \begin{deluxetable*}{lrrrrr}
}{
    \begin{deluxetable}{lrrrrr}
}
\tablewidth{0pc}
\tablecaption{
	Relative radial velocities, bisector spans, and activity index
	measurements of \hatcur{}.
	\label{tab:rvs}
}
\tablehead{
	\colhead{BJD} & 
	\colhead{RV\tablenotemark{a}} & 
	\colhead{\ensuremath{\sigma_{\rm RV}}\tablenotemark{b}} & 
	\colhead{BS} & 
	\colhead{\ensuremath{\sigma_{\rm BS}}} & 
	\colhead{S\tablenotemark{c}} \\%& 
%	\colhead{\ensuremath{\sigma_{\rm S}}}\\
%%
	\colhead{\hbox{(2,454,000$+$)}} & 
	\colhead{(\ms)} & 
	\colhead{(\ms)} &
	\colhead{(\ms)} &
    \colhead{(\ms)} &
	\colhead{} %&
%	\colhead{}
}
\startdata
%%
%% If the table is too long, then we give only sample lines in the
%% submitted version, and the full table is presented in the electronic
%% version.  As regards the astroph version: in such cases we can
%% decide whether to include all data.
%%
\ifthenelse{\boolean{rvtablelong}}{\input{rvtable.tex}
	[-2ex]
}{\input{rvtable_short.tex}
	[-2ex]
}
\enddata
\tablenotetext{a}{
	The zero-point of these velocities is arbitrary. An overall
        offset $\gamma_{\rm rel}$ fitted to these velocities in
        \refsecl{globmod} has {\em not} been subtracted.
}
\tablenotetext{b}{
	Internal errors excluding the component of astrophysical
        jitter considered in \refsecl{globmod}.
}
\tablenotetext{c}{
	Relative chromospheric activity index, not calibrated to the
	scale of \citet{vaughan:1978}.
}
\ifthenelse{\boolean{rvtablelong}}{
	\tablecomments{
		Note that for the iodine-free template exposures we do not
		measure the RV but do measure the BS and S index. Such template
		exposures can be distinguished by the missing RV value. 
	}
}{
    \tablecomments{
		Note that for the iodine-free template exposures we do not
		measure the RV but do measure the BS and S index. Such template
		exposures can be distinguished by the missing RV value. 
        This table is presented in its entirety in the electronic edition
        of the Astrophysical Journal. A portion is shown here for guidance
        regarding its form and content.
	}
} 
\ifthenelse{\boolean{emulateapj}}{
    \end{deluxetable*}
}{
    \end{deluxetable}
}
%% --------------------------------------------------------------------

% =====================================================================
\subsection{Photometric follow-up observations}
\label{sec:phot}

%% --------------------------------------------------------------------
\begin{figure}[!ht]
\plotone{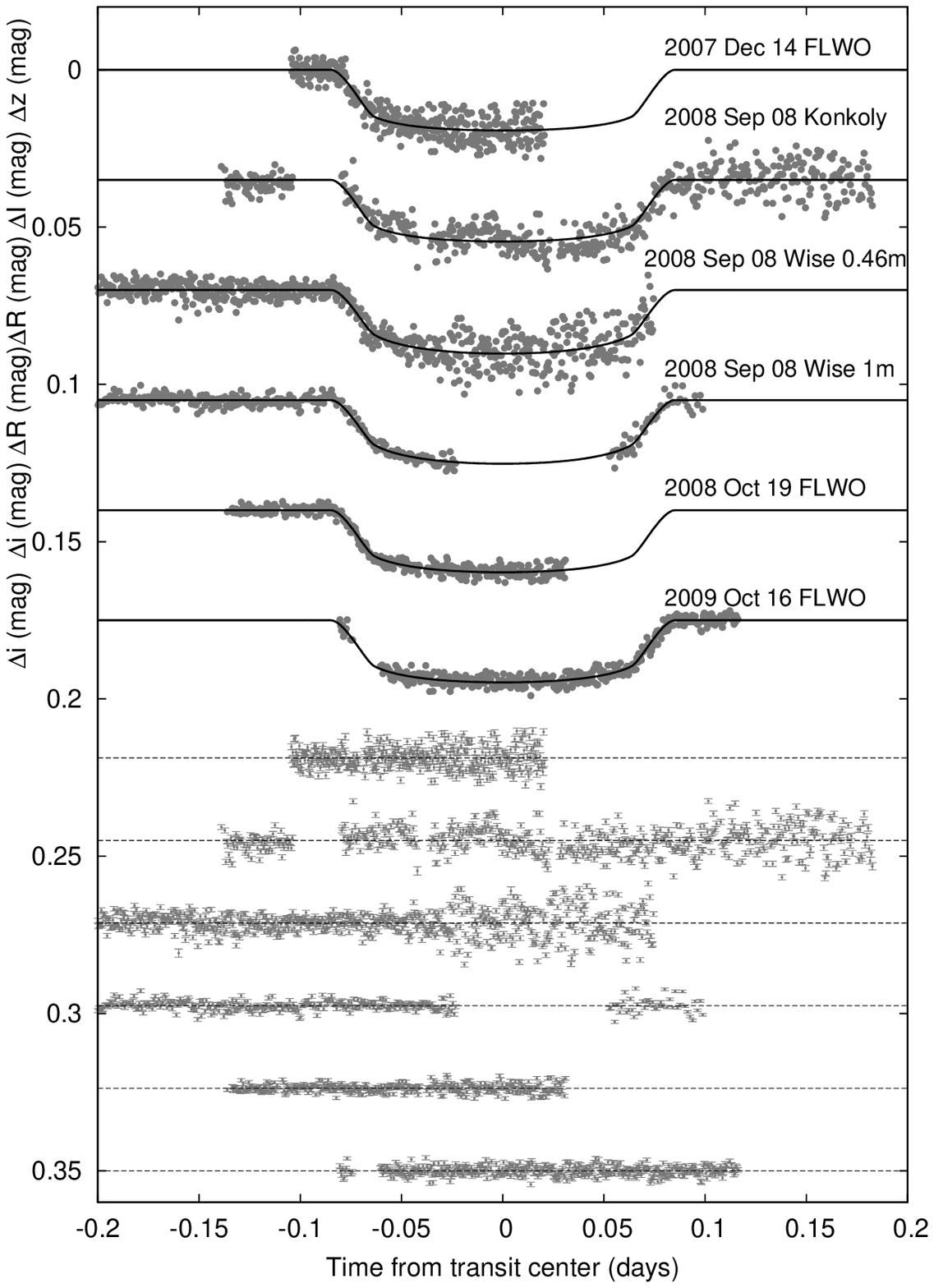}
\caption{
	Unbinned instrumental $z$-band, $I$-band, $R$-band, and
        $i$-band transit \lcs{}, acquired with KeplerCam at the
        \flwof{} telescope, with the Schmidt telescope at Konkoly
        Observatory, and with the 0.46\,m and 1\,m telescopes at Wise
        Observatory. The light curves have been EPD- and TFA-processed,
        as described in \refsec{globmod}.
    The dates and facilities used to observe the events are indicated.
    Curves after the first are displaced vertically for clarity.  Our
    best fit from the global modeling described in \refsecl{globmod}
    is shown by the solid lines.  Residuals from the fits are
    displayed at the bottom, in the same order as the top curves.  The
    error bars represent photon and background shot noise, plus
    readout noise.
\label{fig:lc}}
\end{figure}
%% --------------------------------------------------------------------

%% Observations
%%
In order to permit a more accurate modeling of the light curve, we
conducted additional photometric observations using a variety of
facilities, including: the KeplerCam CCD camera on the \flwof{}
telescope, the 0.6\,m Schmidt telescope of Konkoly Observatory at the
Piszk\'estet\H{o} Mountain Station, and the 0.46\,m and 1.0\,m
telescopes at Wise Observatory. We observed three transit events of
\hatcur{} with the \flwof{} telescope on the nights of 2007 December
14, 2008 October 19, and 2009 October 16, while the event on 2008
September 8 was observed simultaneously at Konkoly Observatory and
with the two telescopes at Wise Observatory (\reffigl{lc}). These
observations are summarized in \reftabl{phfusummary}.

%% --------------------------------------------------------------------
%% Table summarizing photometric Follow-up observations
%%
\ifthenelse{\boolean{emulateapj}}{
    \begin{deluxetable*}{llrrr}
}{
    \begin{deluxetable}{llrrr}
}
\tablewidth{0pc}
\tabletypesize{\scriptsize}
\tablecaption{
    Summary of photometric follow-up observations
    \label{tab:phfusummary}
}
\tablehead{
    \colhead{Facility}  &
    \colhead{Date} &
    \colhead{Number of Images} &
    \colhead{Cadence (s)} &
    \colhead{Filter}
}
\startdata
KeplerCam/\flwof{} & 2007 Dec 14 & 367 & 30 & Sloan \band{z} \\
Konkoly Schmidt 0.6\,m & 2008 Sep 8 & 538 & 45 & \band{I} \\
Wise 0.46\,m & 2008 Sep 8 & 769 & 35 & \band{R} \\
Wise 1.0\,m & 2008 Sep 8 & 407 & 50 & \band{R} \\
KeplerCam/\flwof{} & 2008 Oct 19 & 350 & 33 & Sloan \band{i} \\
KeplerCam/\flwof{} & 2009 Oct 16 & 403 & 33 & Sloan \band{i} \\
[-2ex]
\enddata
\ifthenelse{\boolean{emulateapj}}{
    \end{deluxetable*}
}{
    \end{deluxetable}
}
%% --------------------------------------------------------------------

%% Reduction of frames.
The reduction of these images, including basic calibration,
astrometry, and aperture photometry, was performed as described by
\citet{bakos:2009}. We performed EPD and TFA to remove trends
simultaneously with the light curve modeling
(see \refsecl{analysis}, and \citet{bakos:2009} for details). 
The final time series are
shown in the top portion of \reffigl{lc}, along with our best-fit
transit \lc{} model described below; the individual measurements are
reported in \reftabl{phfu}.

%% --------------------------------------------------------------------
\begin{deluxetable}{lrrrr}
\tablewidth{0pc}
\tablecaption{High-precision differential photometry of \hatcur\label{tab:phfu}}
\tablehead{
	\colhead{BJD} & 
	\colhead{Mag\tablenotemark{a}} & 
	\colhead{\ensuremath{\sigma_{\rm Mag}}} &
	\colhead{Mag(orig)\tablenotemark{b}} & 
	\colhead{Filter} \\
	\colhead{\hbox{~~~~(2,400,000$+$)~~~~}} & 
	\colhead{} & 
	\colhead{} &
	\colhead{} & 
	\colhead{}
}
\startdata
\input{phfu_tab_short.tex}
[-2ex]
\enddata
\tablenotetext{a}{
	The out-of-transit level has been subtracted. These magnitudes have
	been subjected to the EPD and TFA procedures, carried out
	simultaneously with the transit fit.
}
\tablenotetext{b}{
	Raw magnitude values without application of the EPD
	and TFA procedures.
}
\tablecomments{
    This table is available in a machine-readable form in the
    online journal. A portion is shown here for guidance regarding
    its form and content.
}
\end{deluxetable}
%% --------------------------------------------------------------------

% #####################################################################
%% Analysis
\section{Analysis}
\label{sec:analysis}

% =====================================================================
\subsection{Properties of the parent star}
\label{sec:stelparam}

%% Stellar atmosphere parameters
%%
Fundamental parameters of the host star \hatcur{} such as the mass
(\mstar) and radius (\rstar), which are needed to infer the planetary
properties, depend strongly on other stellar quantities that can be
derived spectroscopically.  For this we have relied on the HIRES template
spectrum, and the analysis
package known as Spectroscopy Made Easy \citep[SME;][]{valenti:1996},
along with the atomic line database of \cite{valenti:2005}.  SME
yielded the following {\em initial} values and uncertainties (which we
have conservatively increased to include our estimates of the
systematic errors):
effective temperature $\teffstar=\hatcurSMEiteff$\,K, 
stellar surface gravity $\loggstar=\hatcurSMEilogg$\,(cgs),
metallicity $\feh=\hatcurSMEizfeh$\,dex, and 
projected rotational velocity $\vsini=\hatcurSMEivsin\,\kms$.
We adopt the single-sided uncertainty of $\pm0.5\,\kms$ on \vsini\ 
from \cite{valenti:2005} based on their SME analysis of 
nearly 2000 stars.  
For this star and others with low \vsini, 
the true error distribution excludes unphysical values 
($\vsini < 0\,\kms$) and is likely asymmetric.

In principle the effective temperature and metallicity, along with the
surface gravity taken as a luminosity indicator, could be used as
constraints to infer the stellar mass and radius by comparison with
stellar evolution models.  However, the effect of \loggstar\ on the
spectral line shapes is subtle, and as a result it is typically
difficult to determine accurately, so that  in practice it is a poor
luminosity indicator.  For planetary transits, a stronger
constraint is often provided by \arstar, the normalized semimajor
axis, which is closely related to \rhostar, the mean stellar density.
The quantity \arstar\ can be derived directly from the transit
\lcs\ \citep[see][and also \refsecl{globmod}]{sozzetti:2007}. This, in
turn, improves our determination of the spectroscopic
parameters by supplying an indirect constraint on the weakly
determined spectroscopic value of \loggstar\ that removes
degeneracies. We take this approach here, as described below. The
validity of our assumption, namely that the adequate physical model
describing our data is a planetary transit (as opposed to a blend), is
shown later in \refsecl{bisec}.

Our initial values of \teffstar, \loggstar, and \feh\ were used to
determine auxiliary quantities needed in the global modeling of the
follow-up photometry and radial velocities (specifically, the
limb-darkening coefficients). This modeling, the details of which are
described in \refsecl{globmod}, uses a Monte Carlo approach to deliver
the probability distribution of \arstar\ and other fitted
variables. See \cite{pal:2009b} for further details.  
When combining \arstar\ (a luminosity proxy)
with assumed Gaussian distributions for \teffstar\ and
\feh\ from SME, 
we compare with stellar evolution models to estimate the 
probability distributions of additional inferred stellar parameters, 
including \loggstar.
Here we use the
stellar evolution calculations from the \hatcurisofull\ series by
\cite{\hatcurisocite}.  The comparison with the model isochrones
was carried out for each of 20,000 Monte Carlo trial sets (see
\refsecl{globmod}). Parameter combinations corresponding to unphysical
locations in the \hbox{H-R} diagram (1.5\% of the trials) were
ignored, and replaced with another randomly drawn parameter set.  The
result and error estimate for the surface gravity, $\loggstar = \hatcurISOlogg$, is
different from the result of our initial SME analysis, which is not
surprising in view of the strong correlations among \teffstar, \feh,
and \loggstar\ that are often present in spectroscopic
determinations. Therefore, we carried out a second iteration in which
we adopted this value of \loggstar\ and held it fixed in a new SME
analysis (coupled with a new global modelling of the RV and \lcs),
adjusting only \teffstar, \feh, and \vsini. This gave
$\teffstar = \hatcurSMEiiteff$\,K, 
$\feh = \hatcurSMEiizfeh$, and 
$\vsini = \hatcurSMEiivsin$\,\kms,
in which the conservative uncertainties for the first two have been
increased by a factor of two over their formal values, as before.  A
further iteration did not change \loggstar\ significantly, so we
adopted the values stated above as the final atmospheric properties of
the star.  They are collected in \reftabl{stellar}, together with the
adopted values for the macroturbulent and microturbulent velocities.

\begin{figure}[!ht]
\plotone{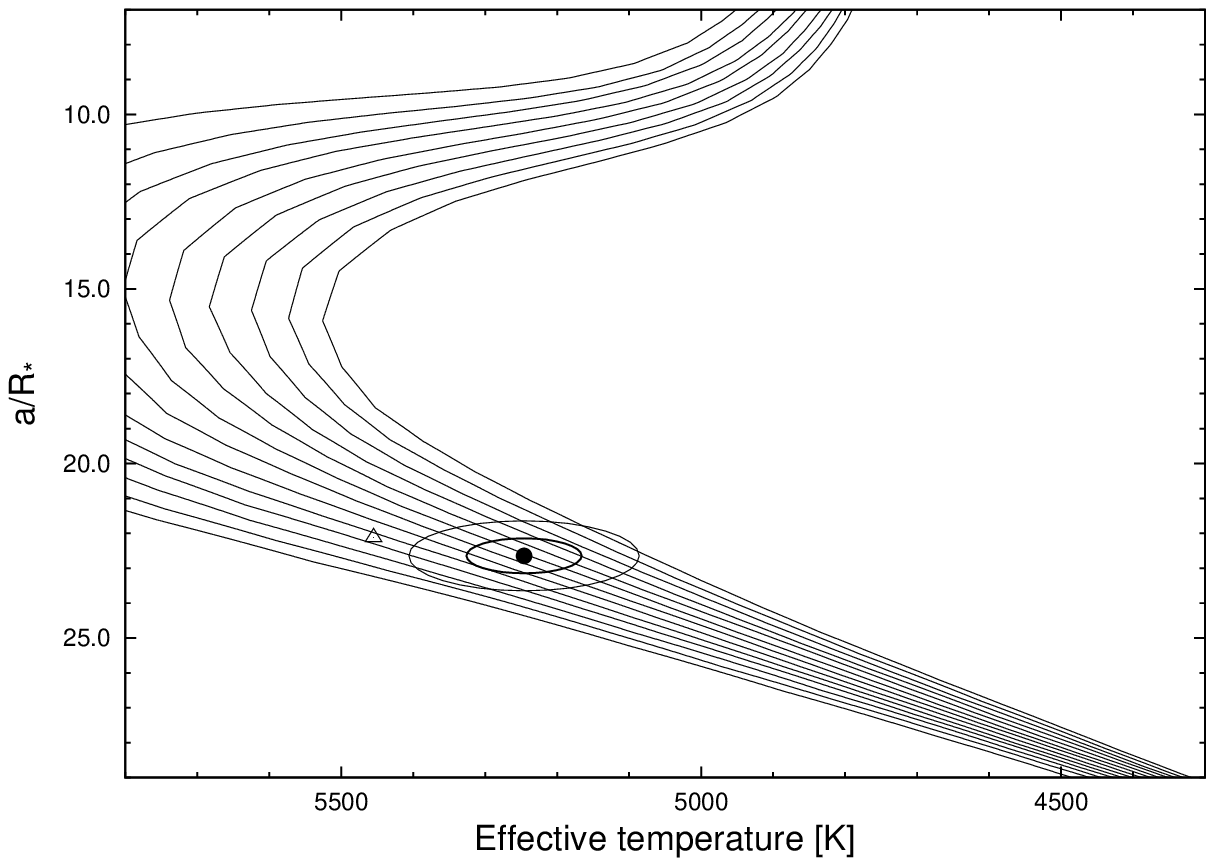}
\caption{
	Model isochrones from \cite{\hatcurisocite} for the measured
        metallicity of \hatcur, \feh = \hatcurSMEiizfehshort, and ages
        between 1.0 and 13.0\,Gyr with a step-size of 1.0\,Gyr (left
        to right).  The adopted values of $\teffstar$ and \arstar\ are
        shown together with their 1$\sigma$ and 2$\sigma$ confidence
        ellipsoids. The initial values of \teffstar\ and \arstar\ from
        the first SME and \lc\ analyses are represented with a
        triangle.
\label{fig:iso}}
\end{figure}

%% Plot of isochrones
%% 
With the adopted spectroscopic parameters the model isochrones yield
the stellar mass and radius, \mstar\ = \hatcurISOmlong\,\msun\ and
\rstar\ = \hatcurISOrlong\,\rsun, along with other properties listed
at the bottom of \reftabl{stellar}. \hatcur{} is a
\hatcurISOspec\ dwarf star with an estimated age of
\hatcurISOage\,Gyr, according to these models.  The inferred location
of the star in a diagram of \arstar\ versus \teffstar, analogous to
the classical H-R diagram, is shown in \reffigl{iso}. The stellar
properties and their 1$\sigma$ and 2$\sigma$ confidence ellipsoids are
displayed against the backdrop of \cite{\hatcurisocite} isochrones for
the measured metallicity of \feh\ = \hatcurSMEiizfehshort, and a range
of ages. For comparison, the location implied by the initial SME
results is also shown (triangle).

%% --------------------------------------------------------------------
%% --------------------------------------------------------------------

%% Distance and colors
%%
The stellar evolution modelling provides color indices that may be
compared against the measured values as a sanity check. The best
available measurements are the near-infrared magnitudes from the 2MASS
Catalogue \citep{skrutskie:2006},
$J_{\rm 2MASS}=\hatcurCCtwomassJmag$, 
$H_{\rm 2MASS}=\hatcurCCtwomassHmag$ and 
$K_{\rm 2MASS}=\hatcurCCtwomassKmag$;
which we have converted to the photometric system of the models (ESO
system) using the transformations by \citet{carpenter:2001}. The
resulting measured color index is $J-K = \hatcurCCesoJKmag$.  This is
within 1$\sigma$ of the predicted value from the isochrones of $J-K =
\hatcurISOJK$. The distance to the object may be computed from the
absolute $K$ magnitude from the models ($M_{\rm K}=\hatcurISOMK$) and
the 2MASS $K_s$ magnitude, which has the advantage of being less
affected by extinction than optical magnitudes.  The result is
$\hatcurXdist$\,pc, where the uncertainty excludes possible
systematics in the model isochrones that are difficult to quantify.

%% --------------------------------------------------------------------
%% Table of stellar parameters. 
%%
\begin{deluxetable}{lcl}
\tablewidth{0pc}
\tabletypesize{\scriptsize}
\tablecaption{
	Stellar parameters for \hatcur{}
	\label{tab:stellar}
}
\tablehead{
	\colhead{~~~~~~~~Parameter~~~~~~~~}	&
	\colhead{Value}                         &
	\colhead{Source}
}
\startdata
\noalign{\vskip -3pt}
\sidehead{Spectroscopic properties}
~~~~$\teffstar$ (K)\dotfill         &  \hatcurSMEteff   & SME\tablenotemark{a}\\
~~~~$\feh$\dotfill                  &  \hatcurSMEzfeh   & SME                 \\
~~~~$\vsini$ (\kms)\dotfill         &  \hatcurSMEvsin   & SME                 \\
~~~~$\vmac$ (\kms)\dotfill          &  \hatcurSMEvmac   & SME                 \\
~~~~$\vmic$ (\kms)\dotfill          &  \hatcurSMEvmic   & SME                 \\
~~~~$\gamma_{\rm RV}$ (\kms)\dotfill &  \hatcurDSgamma   & DS                  \\
\sidehead{Photometric properties}
~~~~$V$ (mag)\dotfill               &  \hatcurCCtassmv  & TASS                \\
~~~~$V\!-\!I_C$ (mag)\dotfill       &  \hatcurCCtassvi  & TASS                \\
~~~~$J$ (mag)\dotfill               &  \hatcurCCtwomassJmag & 2MASS           \\
~~~~$H$ (mag)\dotfill               &  \hatcurCCtwomassHmag & 2MASS           \\
~~~~$K_s$ (mag)\dotfill             &  \hatcurCCtwomassKmag & 2MASS           \\
\sidehead{Derived properties}
~~~~$\mstar$ ($\msun$)\dotfill      &  \hatcurISOmlong   & \hatcurisoshort+\hatcurlumind+SME \tablenotemark{b}\\
~~~~$\rstar$ ($\rsun$)\dotfill      &  \hatcurISOrlong   & \hatcurisoshort+\hatcurlumind+SME         \\
~~~~$\loggstar$ (cgs)\dotfill       &  \hatcurISOlogg    & \hatcurisoshort+\hatcurlumind+SME         \\
~~~~$\lstar$ ($\lsun$)\dotfill      &  \hatcurISOlum     & \hatcurisoshort+\hatcurlumind+SME         \\
~~~~$M_V$ (mag)\dotfill             &  \hatcurISOmv      & \hatcurisoshort+\hatcurlumind+SME         \\
~~~~$M_K$ (mag,\hatcurjhkfilset)\dotfill &  \hatcurISOMK & \hatcurisoshort+\hatcurlumind+SME         \\
~~~~Age (Gyr)\dotfill               &  \hatcurISOage     & \hatcurisoshort+\hatcurlumind+SME         \\
~~~~Distance (pc)\dotfill           &  \hatcurXdist\phn  & \hatcurisoshort+\hatcurlumind+SME\\
[-2ex]
\enddata
\tablenotetext{a}{
	SME = ``Spectroscopy Made Easy'' package for the analysis of
	high-resolution spectra \citep{valenti:1996}. These parameters
	rely primarily on SME, but have a small dependence also on the iterative
	analysis incorporating the isochrone search and global modelling of
	the data, as described in the text.
}
\tablenotetext{b}{
	\hatcurisoshort+\hatcurlumind+SME = Based on the 
        \hatcurisoshort\ isochrones \citep{\hatcurisocite},
        \hatcurlumind\ as a luminosity indicator, and the SME results.
}
\end{deluxetable}
%% --------------------------------------------------------------------

%% Here we can describe rotation and activity. See e.g. HAT-P-11b for
%% details. 

% =====================================================================
\subsection{Spectral line-bisector analysis}
\label{sec:bisec}

Our initial spectroscopic analyses discussed in \refsecl{recspec} and
\refsecl{hispec} rule out the most obvious astrophysical false positive
scenarios.  However, more subtle phenomena such as blends
(contamination by an unresolved eclipsing binary, whether in the
background or associated with the target) can still mimic both the
photometric and spectroscopic signatures we see. 

Following \cite{torres:2007}, we explored the possibility that the
measured radial velocities are not the results of a planet in 
Keplerian motion, but are instead caused by
distortions in the spectral line profiles due to contamination from a
nearby unresolved eclipsing binary. A bisector analysis based on the
Keck spectra was done as described in \S 5 of \cite{bakos:2007a}.  
We detect no variation in excess of the measurement errors in the
bisector spans (see \reffigl{rvbis}). The correlation
between the radial velocities and the bisector variations is
insignificant. Therefore, we conclude that the velocity variations are
real, and that the star is orbited by a close-in giant planet.

% =====================================================================
\subsection{Global modelling of the data}
\label{sec:globmod}

%% Follow-up data
%%
This section describes the procedure we followed to model the HATNet
photometry, the follow-up photometry, and the radial velocities
simultaneously. Our model for the follow-up \lcs\ used analytic
formulae based on \citet{mandel:2002} for the eclipse of a star by a
planet, with limb darkening being prescribed by a quadratic law. The
limb darkening coefficients for the $I$, $R$ and Sloan $z$ and $i$
bands were interpolated from the tables by \citet{claret:2004} for the
spectroscopic parameters of the star as determined from the SME
analysis (\refsecl{stelparam}). The transit shape was parametrized by
the normalized planetary radius $p\equiv \rpl/\rstar$, the square of
the impact parameter $b^2$, and the reciprocal of the half duration of
the transit $\zrstar$. We chose these parameters because of their
simple geometric meanings and their negligible correlations with each 
other \citep[see][]{bakos:2009}. The relation between $\zrstar$
and the quantity \arstar, used in \refsecl{stelparam}, is given by
\begin{equation}
\arstar = P/2\pi (\zrstar) \sqrt{1-b^2} \sqrt{1-e^2}/(1+e \sin\omega)
\end{equation}
\citep[see, e.g.,][]{tingley:2005}. Our model for the HATNet data was
the simplified ``P1P3'' version of the \citet{mandel:2002} analytic
functions (an expansion in terms of Legendre polynomials), for the
reasons described in \citet{bakos:2009}.

%% RV
%%
Initial modelling of the RV observations showed deviations from a
Keplerian fit highly suggestive of a second body in the system with a
much longer period than the transiting planet. Thus, in our global
modelling, the RV curve was parametrized by the combination of an
eccentric Keplerian orbit for the inner planet with semi-amplitude
$K$, and Lagrangian orbital elements $(k,h) \equiv
e\times(\cos\omega,\sin\omega)$, plus an eccentric Keplerian orbit for
the outer object with $K_2$, $k_2$ and $h_2$, and a systemic RV
zero-point $\gamma$ \citep[see also][]{bakos:2009}. Throughout this
paper the subscripts ``1'' and ``2'' refer to \hatcurb\ and
\hatcurc, respectively. If the subscript is omitted, we refer to
\hatcurb.

We assumed a strict periodicity in the individual
transit times. We assigned the transit number $N_{tr} = 0$ to the last
complete follow-up \lc\ gathered on 2009 Oct 16.  The adjustable
parameters in the fit that determine the ephemeris were chosen to be
the time of the first transit center observed with HATNet,
$T_{c,-65}$, and that of the last transit center observed with the
\flwof\ telescope, $T_{c,0}$. We used these as opposed to period and
reference epoch in order to minimize correlations between parameters
\citep[see][]{pal:2008}. Times of mid-transit for intermediate events
were interpolated using these two epochs and the corresponding transit
number of each event, $N_{tr}$.  The eleven main parameters describing
the physical model were thus $T_{c,-65}$, $T_{c,0}$, $\rpl/\rstar$,
$b^2$, $\zrstar$, $K$, $k \equiv e\cos\omega$, $h \equiv e\sin\omega$,
$K_2$, $k_2$ and $h_2$. Five additional parameters were included that
have to do with the instrumental configuration. These are the HATNet
blend factor $B_{\rm inst,247}$, and $B_{\rm inst,248}$ which accounts
for possible dilution of the transit in the \hatcurfield{} and
\hatcurfieldtwo{} HATNet \lcs\ from background stars due to the broad
PSF (20\arcsec\ FWHM), the HATNet out-of-transit magnitudes
$M_{\rm 0,HATNet,247}$, and $M_{\rm 0,HATNet,248}$, and the relative 
zero-point $\gamma_{\rm rel}$ of the Keck RVs.

We extended our physical model with an instrumental model that
describes brightness variations caused by systematic errors in the
measurements. This was done in a similar fashion to the analysis
presented by \citet{bakos:2009}. The HATNet photometry has already
been EPD- and TFA-corrected before the global modeling, so we only
considered corrections for systematics in the follow-up \lcs. We chose
the ``ELTG'' method, i.e., EPD was performed in ``local'' mode with
EPD coefficients defined for each night, and TFA was performed in
``global'' mode using the same set of stars and TFA coefficients for
all nights. 
The total number of fitted
parameters was 16 (physical model with 5
configuration-related parameters) + 36 (local EPD) + 10
(global TFA) = 67, i.e.~much smaller than the number of data points
(2866, counting only RV measurements and follow-up photometry
measurements).

The joint fit was performed as described in \citet{bakos:2009}.  We
minimized \chisq\ in the space of parameters using a hybrid
algorithm, combining the downhill simplex method \citep[AMOEBA;
  see][]{press:1992} with a classical linear least squares algorithm.
Parameter uncertainties were derived applying the Markov
Chain Monte-Carlo method \citep[MCMC, see][]{ford:2006} using
``Hyperplane-CLLS'' chains \citep{bakos:2009}. This provided the full
{\em a posteriori} probability distributions of all adjusted
variables. The {\em a priori} distributions of the parameters for
these chains were chosen to be Gaussian, with eigenvalues and
eigenvectors derived from the Fisher covariance matrix for the
best-fit solution. The Fisher covariance matrix was calculated
analytically using the partial derivatives given by \citet{pal:2009}.

Following this procedure we obtained {\em a posteriori}
distributions for all fitted variables, and other quantities of
interest such as \arstar. As described in \refsecl{stelparam},
\arstar\ was used with stellar evolution models to infer a
theoretical value for \loggstar\ that is significantly more accurate
than the spectroscopic value. The improved estimate was in turn
applied to a second iteration of the SME analysis, as explained
previously, to obtain better estimates of \teffstar\ and
\feh.  The global modeling was then repeated with updated
limb-darkening coefficients based on those new spectroscopic
determinations. The resulting geometric parameters pertaining to the
light curves and velocity curves are listed in
\reftabl{planetparam} and \reftabl{planetparamc}.

Included in \reftabl{planetparam} is the RV ``jitter''. 
This quantity accounts for RV variability due to 
rotational modulation of stellar surface features, 
stellar pulsation, undetected planets, 
and uncorrected systematic errors in the velocity reduction 
\citep{wright:2005}.  
Our adopted jitter value of \hatcurRVjitter\ \ms\ was chosen 
so that reduced $\chi^{2} = 1$ for the RV data of the global
fit.  This value is consistent with the jitter of an ensemble of 
chromospherically quiet, late G/early K dwarf stars 
\citep{wright:2005}.
Auxiliary parameters not listed in the tables are:
$T_{\mathrm{c},-65}=\hatcurLCTA$~(BJD),
$T_{\mathrm{c},0}=\hatcurLCTB$~(BJD), the blending factors 
$B_{\rm instr,247}=\hatcurLCiblendA$ and 
$B_{\rm instr,248}=\hatcurLCiblendB$, and
$\gamma_{\rm rel}=\hatcurRVgamma$\,\ms.
The latter quantity represents an arbitrary offset for the Keck RVs,
and does \emph{not} correspond to the true center of mass velocity of
the system, which was listed earlier in \reftabl{stellar}
($\gamma_{\rm RV}$).

The planetary parameters and their uncertainties are derived
from the {\em a posteriori} distributions of the stellar, \lc,
and RV parameters.  We find an inner planet mass of
$\mpl=\hatcurPPmlong\,\mjup$ and a radius of
$\rpl=\hatcurPPrlong\,\rjup$, giving a mean density
$\rho_p=\hatcurPPrho$\,\gcmc. These and other planetary parameters are
listed at the bottom of Table~\ref{tab:planetparam}.
We note that the inner planets's eccentricity is
significantly non-zero: $e = \hatcurRVeccen$, $\omega =
\hatcurRVomega^\circ$. 

\begin{figure*}[!ht]
\plotone{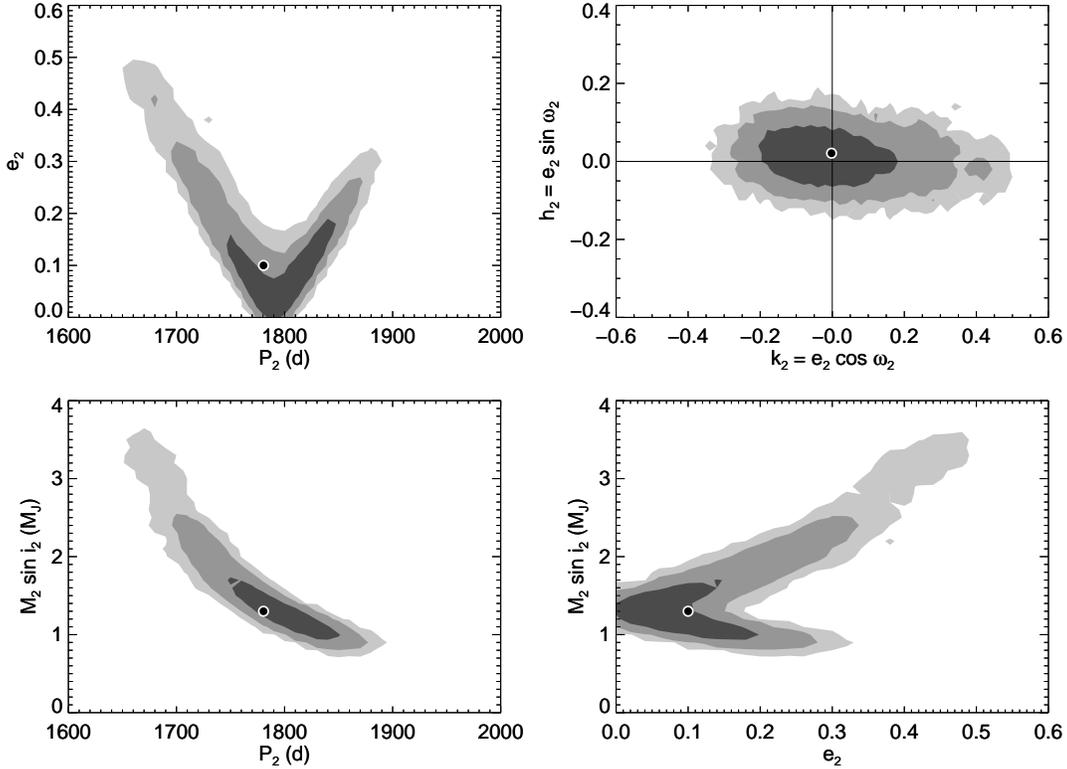}
\caption{
	\textit{A posteriori} distributions showing correlations 
	between parameters describing \hatcurc.  
	Best-fit parameter values are marked with filled circles. 
	Grayscale regions enclose 
	68.3\%, 95.4\%, and 99.73\% of the MCMC samples. 
\label{fig:mcmc}}
\end{figure*}

In addition to \hatcurb, we detect a second, outer planet in the system.  
\hatcurc\ is a long-period jovian planet with a 
minimum mass $m_2 \sin i_2 =
\hatcurcPPmlong\,\mjup$ and 
orbital period $P_2 = \hatcurcLCP$\,days.  
Its eccentricity of $e_2 =
\hatcurcRVeccen$ is consistent with a circular orbit.  
Because we have only measured about half of an orbit of 
\hatcurc, we conservatively adopt 
95.4\% confidence intervals (`2-$\sigma$ errors') 
for the error estimates on parameters associated with this planet.  
(Unless noted, all other parameter uncertainties in this paper are 68.3\% confidence intervals, 
`1-$\sigma$ errors'.)
Figure~\ref{fig:mcmc} shows the distributions of 
and correlations between 
$m_2 \sin i_2$, $P_2$, and $e_2$ from the MCMC analysis.  
Correlations between the Lagrangian orbital parameters 
$k_2 = e_2 \cos \omega_2$ and 
$h_2 = e_2 \sin \omega_2$ are also shown.

%%

%% --------------------------------------------------------------------
%% --------------------------------------------------------------------

% ---------------------------------------------------------------------
\begin{deluxetable}{lc}
%\tablewidth{0pc}
\tabletypesize{\scriptsize}
\tablecaption{Orbital and planetary parameters for \hatcurb{}\label{tab:planetparam}}
\tablehead{
	\colhead{~~~~~~~~~~~~~~~Parameter~~~~~~~~~~~~~~~} &
	\colhead{Value}
}
\startdata
\noalign{\vskip -3pt}
\sidehead{\Lc{} parameters}
~~~$P$ (days)             \dotfill    & $\hatcurLCP$              \\
~~~$T_c$ (${\rm BJD}$)    
      \tablenotemark{a}   \dotfill    & $\hatcurLCT$              \\
~~~$T_{14}$ (days)
      \tablenotemark{a}   \dotfill    & $\hatcurLCdur$            \\
~~~$T_{12} = T_{34}$ (days)
      \tablenotemark{a}   \dotfill    & $\hatcurLCingdur$         \\
~~~$\arstar$              \dotfill    & $\hatcurPPar$             \\
~~~$\zrstar$              \dotfill    & $\hatcurLCzeta$\phn       \\
~~~$\rpl/\rstar$          \dotfill    & $\hatcurLCrprstar$        \\
~~~$b^2$                  \dotfill    & $\hatcurLCbsq$            \\
~~~$b \equiv a \cos i/\rstar$
                          \dotfill    & $\hatcurLCimp$            \\
~~~$i$ (deg)              \dotfill    & $\hatcurPPi$\phn          \\

\sidehead{Limb-darkening coefficients \tablenotemark{b}}
~~~$c_1,i$ (linear term)  \dotfill    & $\hatcurLBii$             \\
~~~$c_2,i$ (quadratic term) \dotfill  & $\hatcurLBiii$            \\
~~~$c_1,z$               \dotfill    & $\hatcurLBiz$             \\
~~~$c_2,z$               \dotfill    & $\hatcurLBiiz$            \\
~~~$c_1,I$               \dotfill    & $\hatcurLBiI$             \\
~~~$c_2,I$               \dotfill    & $\hatcurLBiiI$            \\

\sidehead{RV parameters}
~~~$K$ (\ms)              \dotfill    & $\hatcurRVK$\phn\phn      \\
~~~$k_{\rm RV}$\tablenotemark{c} 
                          \dotfill    & $\hatcurRVk$\phs          \\
~~~$h_{\rm RV}$\tablenotemark{c}
                          \dotfill    & $\hatcurRVh$              \\
~~~$e$                    \dotfill    & $\hatcurRVeccen$          \\
~~~$\omega$ (deg)         \dotfill    & $\hatcurRVomega$\phn      \\
~~~RV jitter (\ms)        \dotfill    & \hatcurRVjitter           \\

\sidehead{Secondary eclipse parameters}
~~~$T_s$ (BJD)            \dotfill    & $\hatcurXsecondary$       \\
~~~$T_{s,14}$              \dotfill    & $\hatcurXsecdur$          \\
~~~$T_{s,12}$              \dotfill    & $\hatcurXsecingdur$       \\

\sidehead{Planetary parameters}
~~~$\mpl$ ($\mjup$)       \dotfill    & $\hatcurPPmlong$          \\
~~~$\rpl$ ($\rjup$)       \dotfill    & $\hatcurPPrlong$          \\
~~~$C(\mpl,\rpl)$
    \tablenotemark{d}     \dotfill    & $\hatcurPPmrcorr$         \\
~~~$\rhopl$ (\gcmc)       \dotfill    & $\hatcurPPrho$            \\
~~~$\log g_p$ (cgs)       \dotfill    & $\hatcurPPlogg$           \\
~~~$a$ (AU)               \dotfill    & $\hatcurPParel$           \\
~~~$T_{\rm eq}$ (K)        \dotfill    & $\hatcurPPteff$           \\
~~~$\Theta$\tablenotemark{e} \dotfill & $\hatcurPPtheta$          \\
~~~$F_{per}$ ($10^{\hatcurPPfluxperidim}$\ergscmsq) \tablenotemark{f}
                          \dotfill    & $\hatcurPPfluxperi$      \\
~~~$F_{ap}$  ($10^{\hatcurPPfluxapdim}$\ergscmsq) \tablenotemark{f} 
                          \dotfill    & $\hatcurPPfluxap$        \\
~~~$\langle F \rangle$ ($10^{\hatcurPPfluxavgdim}$\ergscmsq) \tablenotemark{f}
                          \dotfill    & $\hatcurPPfluxavg$        \\ [-2ex]
\enddata
\tablenotetext{a}{
        \ensuremath{T_c}: Reference epoch of mid transit that
        minimizes the correlation with the orbital period. It
        corresponds to $N_{tr} = -32$.
	\ensuremath{T_{14}}: total transit duration, time
	between first to last contact;
	\ensuremath{T_{12}=T_{34}}: ingress/egress time, time between first
	and second, or third and fourth contact.
}
\tablenotetext{b}{
	Values for a quadratic law, adopted from the tabulations by
        \cite{claret:2004} according to the spectroscopic (SME)
        parameters listed in \reftabl{stellar}.
}
\tablenotetext{c}{
    Lagrangian orbital parameters derived from the global modelling, 
    and primarily determined by the RV data. 
}
\tablenotetext{d}{
	Correlation coefficient between the planetary mass \mpl\ and radius
	\rpl.
}
\tablenotetext{e}{
	The Safronov number is given by $\Theta = \frac{1}{2}(V_{\rm
	esc}/V_{\rm orb})^2 = (a/\rpl)(\mpl / \mstar )$
	\citep[see][]{hansen:2007}.
}
\tablenotetext{f}{
	Incoming flux per unit surface area, averaged over the orbit.
}
\end{deluxetable}

% ---------------------------------------------------------------------
\begin{deluxetable}{lr}
\tablewidth{0pc}
\tablecaption{Orbital and planetary parameters for \hatcurc{}
\label{tab:planetparamc}}
\tablehead{
	\colhead{~~~~~~~~~~~~~~~Parameter~~~~~~~~~~~~~~~} &
	\colhead{Value}
}
\startdata
%%
%% Second planet
%% 
\sidehead{RV parameters, as induced by \hatcurc}
~~~$P_2$ (days)             \dotfill    & $\hatcurcLCP$              \\
~~~$T_{2c}$\tbn{a} (BJD)    \dotfill    & $\hatcurcLCT$              \\
~~~$K_2$ (\ms)              \dotfill    & $\hatcurcRVK$              \\
~~~$k_2$                    \dotfill    & $\hatcurcRVk$              \\
~~~$h_2$                    \dotfill    & $\hatcurcRVh$              \\
~~~$e_2$                    \dotfill    & $\hatcurcRVeccen$          \\
~~~$\omega_2$               \dotfill    & $\hatcurcRVomega^\circ$    \\
~~~$T_{2,peri}$ (days)      \dotfill    & $\hatcurcPPperi$           \\

\sidehead{Hypothetical \lc{} parameters, \hatcurc\tablenotemark{b}}
~~~$T_{2,14}$\tbn{c} (days)   \dotfill    & $\hatcurcLCdur$          \\
~~~$T_{2,12} = T_{34}$ (days) \dotfill    & $\hatcurcLCingdur$       \\

%% Include these only if the eccentricity is non-zero.
\sidehead{Hypothetical secondary eclipse parameters for \hatcurc\tbn{a}}
~~~$T_{2s}$ (BJD)          \dotfill    & $\hatcurcXsecondary$        \\
~~~$T_{2s,14}$ (days)      \dotfill    & $\hatcurcXsecdur$           \\
~~~$T_{2s,12}$ (days)      \dotfill    & $\hatcurcXsecingdur$        \\

\sidehead{Planetary parameters for \hatcurc}
~~~$m_2\sin i_2$ ($\mjup$) \dotfill    & $\hatcurcPPmlong$           \\
~~~$a_2$ (AU)              \dotfill    & $\hatcurcPParel$            \\
~~~$T_{2,\rm eq}$ (K)      \dotfill    & $\hatcurcPPteff$            \\
~~~$F_{2,per}$ ($10^{\hatcurcPPfluxperidim}$\ergscmsq) \tbn{d}
                          \dotfill    & $\hatcurcPPfluxperi$         \\
~~~$F_{2,ap}$  ($10^{\hatcurcPPfluxapdim}$\ergscmsq) \tbn{d} 
                          \dotfill    & $\hatcurcPPfluxap$           \\
~~~$\langle F_2 \rangle$ ($10^{\hatcurcPPfluxavgdim}$\ergscmsq) \tbn{d}
                          \dotfill    & $\hatcurcPPfluxavg$          \\
[-2ex]
\enddata
\tablenotetext{a}{
	$T_{2c}$ would be the center of transit of \hatcurc, if its
	(unknown) inclination is $90\arcdeg$.
}
\tablenotetext{b}{
	Transits of \hatcurc\ have not been observed. The values are for
	guidance only, and assume zero impact parameter.  
}
\tablenotetext{c}{
	\ensuremath{T_{14}}: total transit duration, time
	between first to last contact, assuming zero impact parameter. 
	\ensuremath{T_{12}=T_{34}}: ingress/egress time, time between first
	and second, or third and fourth contact.  Note that these values
	are hypothetical, and transits of \hatcurc\ have not been observed.
}
\tablenotetext{d}{
	Incoming flux per unit surface area in periastron, apastron, and
	averaged over the orbit. 	
}
\end{deluxetable}

% ---------------------------------------------------------------------

%% EOF Analysis

% #####################################################################
%% Discussion
\section{Discussion}
\label{sec:discussion}
	 
We present the detection of \hatcurb, 
a transiting hot Saturn in an eccentric orbit,
and \hatcurc, a cold Jupiter near the ice line 
with an unknown orbital inclination.  
In this section we discuss these two planets in the context of 
recent models and trends, the statistics of nearly 100 TEPs, 
and the small number of multi-planet systems with 
one or more transiting members.

\subsection{The Planet \hatcurb}

\begin{figure*}[!ht]
\plotone{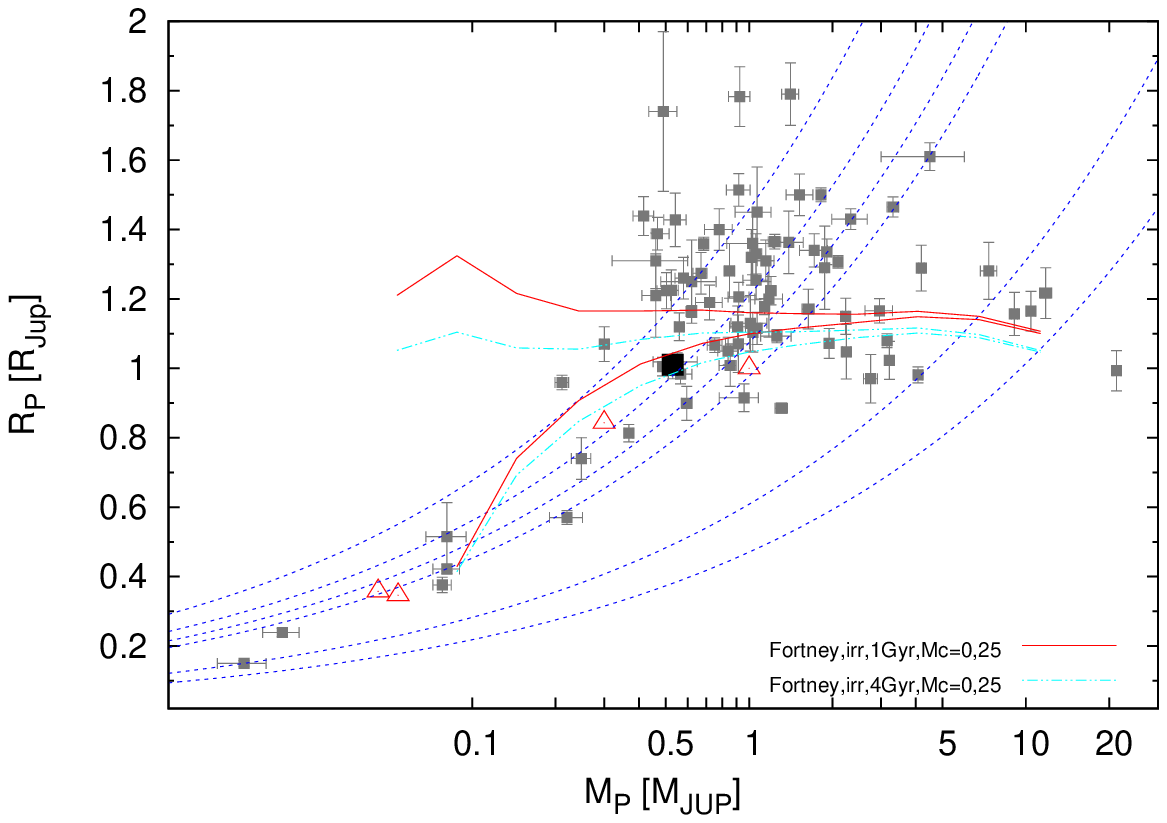}
\caption{
         Mass--radius diagram of known TEPs (small filled squares). 
         \hatcurb\ is shown as a large filled square.
         Overlaid are Fortney et al. (2007) planetary isochrones interpolated 
         to the solar equivalent semi-major axis of \hatcurb\ for ages of 1.0 Gyr 
         (upper solid lines) and 4 Gyr (lower dashed-dotted lines) and 
         core masses of 0 and 25\,\mearth\ (upper and lower lines respectively), 
         as well as isodensity lines for 
         0.4, 0.7, 1.0, 1.33, 5.5 and 11.9\,\gcmc\ (dashed lines). 
         Solar system planets are shown with open triangles.
\label{fig:exomr}}
\end{figure*}

As seen in \reffigl{exomr}, \hatcurb{} has a radius that is 
typical of other known TEPs with masses in the range 
0.5--0.6\,\mjup. Comparing \hatcurb{} to the theoretical 
models by \cite{fortney:2007} we find that it is consistent 
with gas-dominated planet having a core-mass of 
$M_C \sim 25$\,\mearth\ for an age of 4\,Gyr, or somewhat 
less than this for older ages. \hatcurb{} is not inflated relative 
to theoretical models.  
This lack of inflation is consistent with the relatively cool 
temperature ($\teq = \hatcurPPteff$\,K)  of \hatcurb.

%% ----------------
%% ----------------

\subsubsection{A Predicted Non-inverted Atmosphere?}

Secondary eclipse measurements of TEPs in multiple
passbands with the 
\textit{Spitzer Space Telescope} have revealed two 
classes of atmospheres among jovian planets.  
``Non-inverted'' atmospheres are described well 
by 1D atmospheric models with water in absorption.  
In contrast, ``inverted'' atmospheres show 
dayside emission spectra that are best modeled 
by a high-altitude temperature inversion and water 
in emission.  
Physically, such an inversion requires 
a high-altitude absorber.  
Planets in intermediate states (mild inversions, etc) 
also appear possible.  
\cite{knutson:2010} noted that planets with inverted 
atmospheres systematically orbit chromospherically 
quiet stars while planets with non-inverted atmospheres orbit 
chromospherically active stars.  
They proposed that chromospheric activity, 
as measured by the \ion{Ca}{2} H \& K  index 
\logrhk, traces UV flux which is responsible for destroying 
the photochemically labile inversion-causing molecules.  
These authors are agnostic on the exact molecule. 
TiO absorption has been suggested by \cite{hubeny:2003}  
while \cite{zahnle:2009}
argue that sulfur photochemistry is key to generating 
an atmospheric temperature inversion.
\hatcur\ is a chromospherically quiet star with 
$\logrhk = -5.039$.   
Thus, the activity-inversion relation  
predicts that the atmosphere of \hatcurb\ will be inverted.  
The two host stars in the \cite{knutson:2010} 
sample with properties most similar to 
\hatcur---WASP-2 (\teff\ = 5230\,K, $\logrhk = -5.054$) and 
XO-2 (\teff\ = 5340\,K, $\logrhk = -4.988$)---both
have planets with measured temperature inversions.  

A key caveat is that the activity-inversion relation 
is based on a sample of relatively hot planetary atmospheres.  
The photochemistry responsible for atmospheric 
temperature inversion may not be active on
the relatively cool \hatcurb\ (\teq\ = 792\,K).  
For example, the sulfur photochemistry model 
by \cite{zahnle:2009} is only active for $T \ge 1200$\,K.
Observations of this planet in secondary eclipse
will probe the temperature range 
over which such inversions are produced and 
may offer clues to the identity of the absorbing molecules.  

We computed the signal-to-noise ratios (SNR) of 
warm \textit{Spitzer} secondary eclipse observations of \hatcurb\ 
assuming either efficient day-night circulation (\teq\ = 780\,K)
or a hot day side with no circulation (\teq\ = 927\,K).  
For this range of atmospheres, we estimate that \textit{Spitzer} 
will measure the secondary eclipse depth 
with SNR = 2--5 at 3.6\,$\mu$m 
and SNR = 3--6 at 4.5\,$\mu$m for an observation of a single  
secondary eclipse.  
This sensitivity will help constrain atmosphere models for this planet.
The superior collecting area and extended IR coverage of 
\textit{JWST} will give significantly stronger constraints on the 
IR emission spectrum of \hatcurb.

\subsubsection{Spin-orbit Alignment}
 
In the core accretion theory of planet formation, 
hot giant planets like \hatcurb\ form beyond the ice line 
(a few AU from the host star) and subsequently migrate inward.  
Several migration mechanisms have been proposed.  
Tidal interactions with the protoplanetary disk \citep{lin:1996} 
deliver gas giants with uniformly low obliquity.  
Alternatively, Kozai cycles \citep{fabrycky:2007} or 
planet-planet scattering \citep{chatterjee:2009} leave the 
migrated planets in high obliquity orbits, 
possibly with high eccentricity 
(depending on the degree of tidal damping).  
Both high and low-obliquity systems have been observed 
by the Rossiter-McLaughlin (R-M) effect, suggesting some 
combination of migration mechanisms 
(Fabrycky \& Winn 2009; Morton \& Johnson, in preparation).

\citet{winn:2010} recently noted that nearly all misaligned 
(high obliquity) planets orbit hot stars (\teff\ $>$ 6250\,K). 
They suggested that all hot giant planets   
migrated by one of the high obliquity mechanisms and that 
planets orbiting cool stars subsequently 
align the spin axis of the convective zones and photospheres 
of their hosts with the orbital plane.  
Stars above this threshold temperature lack a significant 
convective zone and their close-in giant planets  
remain in high-obliquity orbits.  

Although \hatcur\ is a cool star, the Winn et al.\ 
model predicts spin-orbit misalignment because the 
wider, eccentric orbit of \hatcurb\ lengthens the 
timescale for orbital decay considerably.  
Of the known TEPs,
\hatcurb\ has the longest expected timescale except for 
HD~80606b (J.~Winn, personal communication).
In particular, \hatcurb\ has a longer timescale than WASP-8b, 
which is known to be misaligned \citep{queloz:2010}.  
This planet ($P = 8$ d, $e = 0.31$) is broadly similar to
\hatcurb\ and also orbits a relatively cool star ($\teff = 5600\,K$).

Measuring the projected spin-orbit angle $\lambda$ of \hatcurb\ 
is a challenging but plausible proposition with HIRES.  
We estimate a 4--11\ \ms\ amplitude R-M effect for 
\vsini\ = 0.3--0.8 \kms.

\subsubsection{Similarity to HAT-P-15b}

\hatcurb\ is strikingly similar to HAT-P-15b 
\citep{kovacs:2010} in orbital period 
(10.3\,d and 10.9\,d, respectively) and 
eccentricity (0.35 and 0.19, respectively).  
The mass of HAT-P-15b 
is significantly larger though 
(0.5\,\mjup\ and 1.9\,\mjup, respectively).  
These two planets are the only ground-based transit 
discoveries with orbital periods longer than 10\,d.  
(Other transiting planets with $P > 10$\,d
include those detected by space-based transit 
surveys and two planets discovered by RVs 
that were later shown to transit.)  
The bias toward detecting planets with 
short orbital periods with ground-based transit searches  
stems from the observational window function 
of a longitudinally-spaced multi-site telescope 
network \citep{vonbraun:2009}.  
The HAT-South survey will significantly 
improve the detection of longer period transiting planets 
with a 50\% detection rate out to orbital periods of 
12\,d \citep{bakos-hatsouth:2009}.

\subsection{Transit Timing Varations}

The presence of a second detected  planet in the \hatcur\ 
system raises the possibility of transit timing variations 
\citep[TTVs; ][]{holman:2005}. 
However, because \hatcurb\ and \hatcurc\ are widely separated 
($a_2/a_1 \sim 31$) and \hatcurc\ is on a nearly circular orbit, 
the two planets interact very weakly.
The TTVs are expected to be less than 1~s, 
undetectable with current techniques.

\subsection{The Planet \hatcurc}

\hatcurc\ is an approximately Jupiter-mass planet separated 
from its host star by about half the Sun-Jupiter separation. 
Despite having only observed 50\% of the orbit 
of \hatcurc, its orbital parameters are well constrained 
for a long-period planet (\reffigl{mcmc}).  
While we cannot completely rule out a highly eccentric orbit, 
only 3\% of the MCMC samples have $e > 0.3$.

The 2007 Dec 14 light curve (\reffigl{lc}) showing a partial transit 
of \hatcurb\ is during the broad transit window of \hatcurc.  
We interpret the detected transit as due to \hatcurb\ because the 
timing precisely matches the ephemeris derived from other transits.  
We do not detect additional transits (possibly due to \hatcurc) 
in that light curve.   
Because that light curve is the only one taken in \band{z} 
we cannot compare the photometric level of this transit with others 
to see if it was 
taken entirely when \hatcurc\ was in transit.  

Based on the current orbital fit, the next opportunity to search for a
transit of \hatcurc\ is in 2012 Oct.  
The timing is favorable for an observing campaign as 
the star is visible for $\sim$5\,hr per night from mid-Northern latitudes.  
From the ground, a coordinated, multi-site search spanning 
a range of longitudes is likely necessary to rule in or out 
$\sim$1\% deep transits of maximum duration \hatcurcLCdurshort\,d.

\subsection{Planet Multiplicity}

The migration mechanism of hot jovian planets remains a major 
outstanding problem of planet formation and evolution.  
The presence of additional massive planets in a system 
points to migration within the protoplanetary disk, 
while the absence of additional planets suggests a more 
disruptive mechanism such planet-planet scattering 
or the Kozai mechanism.  

\cite{wright:2009} measured the rate of planet multiplicity and 
found that 14\% of exoplanet host stars are multi-planet systems 
and another 14\% show evidence of multiplicity in the 
form of an RV trend.  
Here we compute the fraction of 
stars hosting a ``cool jovian planet'' 
($\msini > 0.2$\,\mjup\ and $a > 0.2$\, AU)
that also host a ``hot jovian planet'' 
($\msini > 0.2$\,\mjup\ and $a < 0.2$\, AU).  
We used the Exoplanet Orbit Database\footnote{http://exoplanets.org} 
of planets with well-defined orbital parameters.  
Of the 375 planet hosts (including \hatcur), 
we find 106 stars that host a hot jovian planet and 
204 stars that host  one or more cool jovian planets.  
Of the latter group, 10 stars (5\%) also host a hot jovian planet.  
Restricting the hot jovian planets to $a < 0.1$\,AU, 6/204 = 3\% 
of stars host both cool and hot jovian planets.  
Note that this selection of planets does not suffer from a significant 
detection bias; the Doppler signal from a hot jovian planet is 
essentially always detectable for systems with a detected 
cool jovian planet.  
While hot jovian planets represent a 
disproportionally large fraction of the known planets due 
to observational selection effects, 
multi-planet systems like \hatcur\ are rare.

%% EOF Discussion

% #####################################################################
%% Acknowledgements
\acknowledgements 

We thank H.~Knutson, J.~Winn, and J.~Wright for helpful conversations.  
HATNet operations have been funded by NASA grants NNG04GN74G,
NNX08AF23G and SAO IR\&D grants. 
A.W.H.\ gratefully acknowledges support from a Townes Post-doctoral Fellowship 
at the U.\,C.\ Berkeley Space Sciences Laboratory.
Work of G.\'A.B.~and J.~Johnson were
supported by the Postdoctoral Fellowship of the NSF Astronomy and
Astrophysics Program (AST-0702843 and AST-0702821, respectively). G.T.\
acknowledges partial support from NASA grant NNX09AF59G. We acknowledge
partial support also from the Kepler Mission under NASA Cooperative
Agreement NCC2-1390 (D.W.L., PI). G.K.~thanks the Hungarian Scientific
Research Foundation (OTKA) for support through grant K-81373. 
T.M.\ acknowledges the Israel Science Foundation (grant 655/07).
This research has made use of Keck telescope time granted through NOAO
and NASA. We thank Ezra Mashal for his help in operating the
Wise HAT telescope over the past years. We thank the TLC project (M.~Holman
and J.~Winn) for swapping time on the 1.2\,m telescope on a short notice.  
This research has made use of the Exoplanet Orbit Database 
and the Exoplanet Data Explorer at exoplanets.org, 
the SIMBAD database (operated at CDS, Strasbourg, France), and 
NASA's Astrophysics Data System Bibliographic Services.
Finally, the authors wish to extend special thanks to those of Hawai`ian ancestry 
on whose sacred mountain of Mauna Kea we are privileged to be guests.  
Without their generous hospitality, the Keck observations presented herein
would not have been possible.

%% EOF Acknowledgements

% #####################################################################
%% Bibliography
\input{biblio.tex}

\end{document}

%% file: newcommand_gen.tex
\newcommand{\forloop}[5][1]%
{%
\setcounter{#2}{#3}%
\ifthenelse{#4}%
	{%
	#5%
	\addtocounter{#2}{#1}%
	\forloop[#1]{#2}{\value{#2}}{#4}{#5}%
	}%
	{%
	}%
}% 

\newcommand{\ctbd}[1]{}

\newcommand{\lc}{light curve}
\newcommand{\lcs}{light curves}
\newcommand{\Lc}{Light curve}

\newcommand{\band}[1]{\ensuremath{#1}-band}

\newcommand{\chisq}{\ensuremath{\chi^2}}

\newcommand{\kms}{\ensuremath{\rm km\,s^{-1}}}
\newcommand{\ms}{\ensuremath{\rm m\,s^{-1}}}

\newcommand{\gcmc}{\ensuremath{\rm g\,cm^{-3}}}
\newcommand{\ergscmsq}{\ensuremath{\rm erg\,s^{-1}\,cm^{-2}}}

\newcommand{\msini}{\ensuremath{m \sin i}}

\newcommand{\teff}{\ensuremath{T_{\rm eff}}}
\newcommand{\teq}{\ensuremath{T_{\rm eq}}}

\newcommand{\vsini}{\ensuremath{v \sin{i}}}
\newcommand{\feh}{\ensuremath{\rm [Fe/H]}}

\newcommand{\vmac}{\ensuremath{v_{\rm mac}}}
\newcommand{\vmic}{\ensuremath{v_{\rm mic}}}
\newcommand{\rhk}{\ensuremath{R^{\prime}_{\rm HK}}}
\newcommand{\logrhk}{\ensuremath{\log\rhk}}
\newcommand{\rsun}{\ensuremath{R_\sun}}
\newcommand{\msun}{\ensuremath{M_\sun}}
\newcommand{\lsun}{\ensuremath{L_\sun}}

\newcommand{\rstar}{\ensuremath{R_\star}}
\newcommand{\mstar}{\ensuremath{M_\star}}
\newcommand{\lstar}{\ensuremath{L_\star}}

\newcommand{\teffstar}{\ensuremath{T_{\rm eff\star}}}
\newcommand{\rhostar}{\ensuremath{\rho_\star}}
\newcommand{\loggstar}{\ensuremath{\log{g_{\star}}}}

\newcommand{\mearth}{\ensuremath{M_\earth}}

\newcommand{\rpl}{\ensuremath{R_{p}}}
\newcommand{\mpl}{\ensuremath{M_{p}}}

\newcommand{\rhopl}{\ensuremath{\rho_{p}}}

\newcommand{\arstar}{\ensuremath{a/\rstar}}
\newcommand{\zrstar}{\ensuremath{\zeta/\rstar}}

\newcommand{\rjup}{\ensuremath{R_{\rm J}}}
\newcommand{\mjup}{\ensuremath{M_{\rm J}}}

\newcommand{\refsec}[1]{\mbox{\S\ \ref{sec:#1}}}

\newcommand{\reffigl}[1]{Figure~\ref{fig:#1}}
\newcommand{\refsecl}[1]{\mbox{Section \ref{sec:#1}}}

\newcommand{\reftabl}[1]{Table~\ref{tab:#1}}

\newcommand{\flwof}{\mbox{FLWO 1.2\,m}}

\newcommand{\flwos}{\mbox{FLWO 1.5\,m}}

\newcommand{\tbn}[1]{\tablenotemark{#1}}

%% file: newcommand_spec.tex
\input{planetinfo.tex}

\newcommand{\hatcur}{HAT-P-17}
\newcommand{\hatcurb}{HAT-P-17b}
\newcommand{\hatcurc}{HAT-P-17c}
\newcommand{\hatcurbc}{HAT-P-17b,c}
\newcommand{\hatcurCCtassvi}{\ensuremath{0.901\pm0.10}}                  % TASS V-I
\newcommand{\hatcurSMEversion}{ii}                                       % Final SME version:i or ii?
\newcommand{\hatcurSMEteff}{\ifthenelse{\equal{\hatcurSMEversion}{i}}{\hatcurSMEiteff}{\hatcurSMEiiteff}}
\newcommand{\hatcurSMEzfeh}{\ifthenelse{\equal{\hatcurSMEversion}{i}}{\hatcurSMEizfeh}{\hatcurSMEiizfeh}}
\newcommand{\hatcurSMEzfehshort}{\ifthenelse{\equal{\hatcurSMEversion}{i}}{\hatcurSMEizfehshort}{\hatcurSMEiizfehshort}}
\newcommand{\hatcurSMElogg}{\ifthenelse{\equal{\hatcurSMEversion}{i}}{\hatcurSMEilogg}{\hatcurSMEiilogg}}
\newcommand{\hatcurSMEvsin}{\ifthenelse{\equal{\hatcurSMEversion}{i}}{\hatcurSMEivsin}{\hatcurSMEiivsin}}
\newcommand{\hatcurSMEvmac}{\ifthenelse{\equal{\hatcurSMEversion}{i}}{\hatcurSMEivmac}{\hatcurSMEiivmac}}
\newcommand{\hatcurSMEvmic}{\ifthenelse{\equal{\hatcurSMEversion}{i}}{\hatcurSMEivmic}{\hatcurSMEiivmic}}

%% file: planetinfo.tex
\newcommand{\hatcurhtr}{HTR248-002}                                    % Original HTR name of target
\newcommand{\hatcurfield}{247}                                         % Original HTR field
\newcommand{\hatcurfieldtwo}{248}                                         % Original HTR field
\newcommand{\hatcurCCra}{\ensuremath{21^{\mathrm h}38^{\mathrm m}08.88{\mathrm s}}}                                  % Right Ascension
\newcommand{\hatcurCCdec}{\ensuremath{+30{\arcdeg}29{\arcmin}19.4{\arcsec}}}                                 % Declination
\newcommand{\hatcurCCtwomass}{2MASS~21380873+3029193}                  % 2MASS identifier
\newcommand{\hatcurCCgsc}{GSC~2717-00417}                              % GSC(1.2) identifier
\newcommand{\hatcurCCtassmv}{10.54}                                   % TASS V-band magnitude - 10.535
\newcommand{\hatcurCCtwomassJmag}{\ensuremath{9.017\pm0.022}}          % 2MASS ORIG MAG
\newcommand{\hatcurCCtwomassHmag}{\ensuremath{8.619\pm0.029}}          % 2MASS ORIG MAG
\newcommand{\hatcurCCtwomassKmag}{\ensuremath{8.544\pm0.025}}          % 2MASS ORIG MAG
\newcommand{\hatcurCCesoJKmag}{\ensuremath{0.504\pm0.036}}             % 2MASS ESO JK COLOR
\newcommand{\hatcurLCdip}{\ensuremath{12.0}}                           % BLS detected dip (mmag)
\newcommand{\hatcurLCrprstar}{\ensuremath{0.1238\pm0.0009}}            % Rp/R*
\newcommand{\hatcurLCbsq}{\ensuremath{0.097_{-0.034}^{+0.032}}}        % impact parameter square
\newcommand{\hatcurLCimp}{\ensuremath{0.311_{-0.067}^{+0.046}}}        % impact parameter
\newcommand{\hatcurLCzeta}{\ensuremath{13.44\pm0.04}}                  % zeta/R*
\newcommand{\hatcurLCdur}{\ensuremath{0.1691\pm0.0009}}                % transit duration (days)
\newcommand{\hatcurLCdurhr}{\ensuremath{4.058\pm0.022}}                % transit duration (hours)
\newcommand{\hatcurLCq}{\ensuremath{0.0164\pm0.0001}}                  % fractional transit duration (days)
\newcommand{\hatcurLCingdur}{\ensuremath{0.0204\pm0.0009}}             % ingress/egress duration (days)
\newcommand{\hatcurLCP}{\ensuremath{10.338523\pm0.000009}}             % period (days)
\newcommand{\hatcurLCPprec}{\ensuremath{10.3385228}}                   % period (days)
\newcommand{\hatcurLCPshort}{\ensuremath{10.3385}}                     % period (days)
\newcommand{\hatcurLCT}{\ensuremath{2454801.16945\pm0.00020}}          % epoch (BJD)
\newcommand{\hatcurLCTA}{\ensuremath{2454449.65967\pm0.00035}}         % TA (BJD)
\newcommand{\hatcurLCTB}{\ensuremath{2455121.66365\pm0.00034}}         % TB (BJD)
\newcommand{\hatcurLCiblendA}{\ensuremath{0.87\pm0.04}}                % HATNet iblend factor
\newcommand{\hatcurLCiblendB}{\ensuremath{0.69\pm0.16}}                % HATNet iblend factor
\newcommand{\hatcurSMEiteff}{\ensuremath{5455\pm80}}                   % Ini SME, stellar effective temperature
\newcommand{\hatcurSMEizfeh}{\ensuremath{+0.13\pm0.08}}                 % Ini SME, stellar metallicity
\newcommand{\hatcurSMEizfehshort}{\ensuremath{+0.13}}                   % Ini SME, stellar metallicity
\newcommand{\hatcurSMEilogg}{\ensuremath{4.60\pm0.10}}                  % Ini SME, stellar surface gravity
\newcommand{\hatcurSMEivsin}{\ensuremath{0.5\pm0.5}}                   % Ini SME, stellar rotational velocity
\newcommand{\hatcurSMEivmac}{\ensuremath{0.0}}                         % Ini SME, stellar macroturbulence
\newcommand{\hatcurSMEivmic}{\ensuremath{0.0}}                         % Ini SME, stellar microturbulence
\newcommand{\hatcurSMEiiteff}{\ensuremath{5246\pm80}}                  % Final SME, stellar effective temperature
\newcommand{\hatcurSMEiizfeh}{\ensuremath{0.00\pm0.08}}                % Final SME, stellar metallicity
\newcommand{\hatcurSMEiizfehshort}{\ensuremath{0.00}}                  % Final SME, stellar metallicity
\newcommand{\hatcurSMEiilogg}{\ensuremath{4.52\pm0.06}}                % Final SME, stellar surface gravity
\newcommand{\hatcurSMEiivsin}{\ensuremath{0.3\pm0.5}}                  % Final SME, stellar rotational velocity
\newcommand{\hatcurSMEiivmac}{\ensuremath{3.21}}                       % Final SME, stellar macroturbulence
\newcommand{\hatcurSMEiivmic}{\ensuremath{0.85}}                       % Final SME, stellar microturbulence
\newcommand{\hatcurDSteff}{\ensuremath{5250\pm125}}                    % DS stellar effective temperature
\newcommand{\hatcurDSlogg}{\ensuremath{4.5\pm0.25}}                    % DS stellar surface gravity
\newcommand{\hatcurDSvsini}{\ensuremath{0.0^{+0.5}_{-0.0}}}                    % DS stellar rotational velocity
\newcommand{\hatcurDSgamma}{\ensuremath{20.13\pm0.21}}                 % DS absolute gamma velocity
\newcommand{\hatcurDSrvrms}{\ensuremath{0.60}}                         % DS rms of RV values [km/s]
\newcommand{\hatcurLBiz}{\ensuremath{0.2843}}                          % Limb darkening parameters, Gamma1, z-band
\newcommand{\hatcurLBiiz}{\ensuremath{0.2882}}                         % Limb darkening parameters, Gamma2, z-band
\newcommand{\hatcurLBii}{\ensuremath{0.3592}}                          % Limb darkening parameters, Gamma1, i-band
\newcommand{\hatcurLBiii}{\ensuremath{0.2759}}                         % Limb darkening parameters, Gamma2, i-band
\newcommand{\hatcurLBiI}{\ensuremath{0.3347}}                          % Limb darkening parameters, Gamma1, I-band
\newcommand{\hatcurLBiiI}{\ensuremath{0.2800}}                         % Limb darkening parameters, Gamma2, I-band
\newcommand{\hatcurISOm}{\ensuremath{0.86\pm0.04}}                     % stellar mass
\newcommand{\hatcurISOmlong}{\ensuremath{0.857\pm0.039}}               % stellar mass
\newcommand{\hatcurISOr}{\ensuremath{0.84\pm0.02}}                     % stellar radius
\newcommand{\hatcurISOrlong}{\ensuremath{0.837\pm0.021}}               % stellar radius
\newcommand{\hatcurISOlogg}{\ensuremath{4.53\pm0.02}}                  % stellar surface gravity from isochrones
\newcommand{\hatcurISOlum}{\ensuremath{0.48\pm0.04}}                   % stellar luminosity
\newcommand{\hatcurISOmv}{\ensuremath{5.75\pm0.12}}                    % stellar absolute magnitude
\newcommand{\hatcurISOage}{\ensuremath{7.8\pm3.3}}                     % stellar age
\newcommand{\hatcurISOMK}{\ensuremath{3.79\pm0.07}}                    % stellar absolute K magnitude
\newcommand{\hatcurISOJK}{\ensuremath{0.53\pm0.02}}                    % J-K color index from isochrones.
\newcommand{\hatcurISOspec}{early K}                                   % stellar spectral type
\newcommand{\hatcurRVK}{\ensuremath{58.4\pm0.9}}                       % RV semi-amplitude [m/s]
\newcommand{\hatcurRVk}{\ensuremath{-0.321\pm0.006}}                   % e*cos(omega)
\newcommand{\hatcurRVh}{\ensuremath{-0.129\pm0.013}}                   % e*sin(omega)
\newcommand{\hatcurRVgamma}{\ensuremath{25.0\pm7.6}}                   % RV gamma velocity, relative scale
\newcommand{\hatcurRVjitter}{\ensuremath{2.3}}                         % RV jitter (m/s)
\newcommand{\hatcurRVeccen}{\ensuremath{0.346\pm0.007}}                % eccentricity
\newcommand{\hatcurRVomega}{\ensuremath{201\pm2}}                      % argument of pericenter
\newcommand{\hatcurPPi}{\ensuremath{89.2_{-0.1}^{+0.2}}}               % orbital inclination
\newcommand{\hatcurPPlogg}{\ensuremath{3.11\pm0.02}}                   % planetary surface gravity (log cgs)
\newcommand{\hatcurPPar}{\ensuremath{22.65\pm0.50}}                    % relative orbital radius (a/R*)
\newcommand{\hatcurPParel}{\ensuremath{0.0882\pm0.0014}}               % semimajor axis (AU)
\newcommand{\hatcurPPrho}{\ensuremath{0.64\pm0.05}}                    % planetary density (cgs)
\newcommand{\hatcurPPmlong}{\ensuremath{0.530\pm0.018}}                % planetary mass (M_jup)
\newcommand{\hatcurPPrlong}{\ensuremath{1.010\pm0.029}}                % planetary radius (R_jup)
\newcommand{\hatcurPPmrcorr}{\ensuremath{0.44}}                        % mass/radius correlation
\newcommand{\hatcurPPteff}{\ensuremath{792\pm15}}                      % planetary temperature (K)
\newcommand{\hatcurPPtheta}{\ensuremath{0.108\pm0.004}}                % Safranov number
\newcommand{\hatcurPPfluxperi}{\ensuremath{1.95\pm0.15}}               % flux @ periastron (CGS)
\newcommand{\hatcurPPfluxperidim}{\ensuremath{8}}                      % flux @ periastron (CGS) units.
\newcommand{\hatcurPPfluxap}{\ensuremath{4.60\pm0.36}}                 % flux @ apastron (CGS)
\newcommand{\hatcurPPfluxapdim}{\ensuremath{7}}                        % flux @ apastron (CGS) units.
\newcommand{\hatcurPPfluxavg}{\ensuremath{8.89\pm0.66}}                % flux on average (CGS)
\newcommand{\hatcurPPfluxavgdim}{\ensuremath{7}}                       % flux average (CGS) units.
\newcommand{\hatcurXsecondary}{\ensuremath{2454804.246\pm0.037}}       % Secondary eclipse epoch
\newcommand{\hatcurXsecdur}{\ensuremath{0.1326\pm0.0036}}              % sec eclipse duration (days)
\newcommand{\hatcurXsecingdur}{\ensuremath{0.0154\pm0.0007}}           % sec I/E duration (days)
\newcommand{\hatcurcLCdur}{\ensuremath{0.883\pm0.092}}                 % transit duration (days)
\newcommand{\hatcurcLCdurshort}{\ensuremath{0.883}}                    % transit duration (days)
\newcommand{\hatcurcLCingdur}{\ensuremath{0.1008\pm0.0114}}            % ingress/egress duration (days)
\newcommand{\hatcurcLCP}{\ensuremath{1798_{-89}^{+58}}}                   % period (days)
\newcommand{\hatcurcLCT}{\ensuremath{2452636\pm58}}                % epoch (BJD)
\newcommand{\hatcurcRVK}{\ensuremath{25_{-6}^{+24}}}                          % RV semi-amplitude [m/s]
\newcommand{\hatcurcRVk}{\ensuremath{0.00_{-0.22}^{+0.32}}}        % e*cos(omega)
\newcommand{\hatcurcRVh}{\ensuremath{0.02_{-0.09}^{+0.10}}}                   % e*sin(omega)
\newcommand{\hatcurcRVeccen}{\ensuremath{0.10_{-0.10}^{+0.23}}}               % eccentricity
\newcommand{\hatcurcRVomega}{\ensuremath{155\pm150}}               % argument of pericenter
\newcommand{\hatcurcPParel}{\ensuremath{2.75\pm0.12}}                % semimajor axis (AU)
\newcommand{\hatcurcPPmlong}{\ensuremath{1.4_{-0.4}^{+1.1}}}           % planetary mass (M_jup)
\newcommand{\hatcurcPPteff}{\ensuremath{140\pm6}}                      % planetary temperature (K)
\newcommand{\hatcurcPPfluxperi}{\ensuremath{1.1\pm0.5}}             % flux @ periastron (CGS)
\newcommand{\hatcurcPPfluxperidim}{\ensuremath{5}}                     % flux @ periastron (CGS) units.
\newcommand{\hatcurcPPfluxap}{\ensuremath{7.0\pm1.8}}                   % flux @ apastron (CGS)
\newcommand{\hatcurcPPfluxapdim}{\ensuremath{4}}                       % flux @ apastron (CGS) units.
\newcommand{\hatcurcPPfluxavg}{\ensuremath{8.7\pm1.2}}              % flux on average (CGS)
\newcommand{\hatcurcPPfluxavgdim}{\ensuremath{4}}                      % flux average (CGS) units.
\newcommand{\hatcurcXsecondary}{\ensuremath{2453531\pm208}}        % Secondary eclipse epoch
\newcommand{\hatcurcXsecdur}{\ensuremath{0.92\pm0.11}}               % sec eclipse duration (days)
\newcommand{\hatcurcXsecingdur}{\ensuremath{0.101\pm0.011}}          % sec I/E duration (days)
\newcommand{\hatcurcPPperi}{\ensuremath{2452670\pm702}}          % time of periastron passage.
\newcommand{\hatcurXdist}{\ensuremath{90\pm3}}                         % distance (pc)

%% file: rvtable.tex
$ 396.82931 $ \dotfill & $    18.31 $ & $     1.62 $ & $    10.31 $ & $     4.38 $ & $    0.0048 $   \\
$ 397.79661 $ \dotfill & $    -8.73 $ & $     1.63 $ & $     2.84 $ & $     5.15 $ & $    0.0048 $   \\
$ 427.78145 $ \dotfill & $    13.43 $ & $     1.45 $ & $     2.49 $ & $     5.17 $ & $    0.0047 $   \\
$ 427.78895 $ \dotfill & \nodata      & \nodata      & $    -2.14 $ & $     5.34 $ & $    0.0047 $   \\
$ 429.81978 $ \dotfill & $   -46.29 $ & $     1.62 $ & $    -3.08 $ & $     5.46 $ & $    0.0047 $   \\
$ 430.84830 $ \dotfill & $   -74.61 $ & $     1.81 $ & $    -1.88 $ & $     5.52 $ & $    0.0046 $   \\
$ 454.71418 $ \dotfill & $    33.26 $ & $     2.56 $ & $   -12.71 $ & $     6.37 $ & $    0.0048 $   \\
$ 455.70568 $ \dotfill & $    38.42 $ & $     1.72 $ & $     0.09 $ & $     5.13 $ & $    0.0047 $   \\
$ 456.69684 $ \dotfill & $    37.25 $ & $     1.84 $ & $     4.61 $ & $     4.86 $ & $    0.0048 $   \\
$ 457.69351 $ \dotfill & $    24.99 $ & $     1.63 $ & $    -3.52 $ & $     5.47 $ & $    0.0048 $   \\
$ 603.04188 $ \dotfill & $     6.53 $ & $     1.63 $ & $     3.60 $ & $     4.80 $ & $    0.0047 $   \\
$ 604.01999 $ \dotfill & $   -10.16 $ & $     1.63 $ & $    -2.39 $ & $     5.19 $ & $    0.0048 $   \\
$ 638.00747 $ \dotfill & $   -80.35 $ & $     1.56 $ & $     6.54 $ & $     4.54 $ & $    0.0049 $   \\
$ 640.09696 $ \dotfill & $    13.55 $ & $     1.65 $ & $     2.73 $ & $     5.17 $ & $    0.0048 $   \\
$ 641.03845 $ \dotfill & $    21.59 $ & $     2.05 $ & $     9.89 $ & $     4.28 $ & $    0.0049 $   \\
$ 674.90104 $ \dotfill & $     7.88 $ & $     1.90 $ & $   -15.07 $ & $     6.42 $ & $    0.0049 $   \\
$ 722.81909 $ \dotfill & $     5.61 $ & $     1.76 $ & $     9.33 $ & $     4.88 $ & $    0.0048 $   \\
$ 726.78740 $ \dotfill & $     6.77 $ & $     1.52 $ & $    -8.43 $ & $     5.62 $ & $    0.0047 $   \\
$ 727.86651 $ \dotfill & $   -13.81 $ & $     1.89 $ & $    -5.89 $ & $     5.34 $ & $    0.0044 $   \\
$ 777.87074 $ \dotfill & $    11.74 $ & $     1.61 $ & $     8.04 $ & $     4.51 $ & $    0.0048 $   \\
$ 778.81597 $ \dotfill & $     2.41 $ & $     1.95 $ & $     3.34 $ & $     4.82 $ & $    0.0049 $   \\
$ 805.73023 $ \dotfill & $     3.20 $ & $     1.52 $ & $     9.31 $ & $     4.65 $ & $    0.0045 $   \\
$ 808.77451 $ \dotfill & $     6.03 $ & $     1.43 $ & $    -3.86 $ & $     5.37 $ & $    0.0047 $   \\
$ 955.07012 $ \dotfill & $    -7.96 $ & $     1.44 $ & $     5.73 $ & $     4.54 $ & $    0.0048 $   \\
$ 956.08553 $ \dotfill & $   -34.98 $ & $     1.42 $ & $     1.12 $ & $     5.07 $ & $    0.0049 $   \\
$ 957.07704 $ \dotfill & $   -67.81 $ & $     1.91 $ & $    -6.36 $ & $     5.74 $ & $    0.0048 $   \\
$ 985.08988 $ \dotfill & $     0.05 $ & $     1.74 $ & $     1.03 $ & $     5.04 $ & $    0.0048 $   \\
$ 989.04979 $ \dotfill & $  -103.08 $ & $     1.90 $ & $    -0.24 $ & $     5.13 $ & $    0.0047 $   \\
$ 1043.02166 $ \dotfill & $    -1.01 $ & $     1.89 $ & $     9.50 $ & $     4.57 $ & $    0.0048 $   \\
$ 1106.96789 $ \dotfill & $    32.17 $ & $     2.59 $ & $    -1.47 $ & $     5.67 $ & $    0.0048 $   \\
$ 1192.72802 $ \dotfill & $     3.26 $ & $     1.59 $ & $    -6.09 $ & $     5.60 $ & $    0.0050 $   \\
$ 1198.70293 $ \dotfill & $    25.65 $ & $     1.70 $ & $     2.47 $ & $     5.18 $ & $    0.0046 $   \\
$ 1290.13821 $ \dotfill & $   -22.07 $ & $     2.12 $ & $   -21.11 $ & $     7.04 $ & $    0.0045 $   \\
$ 1319.05702 $ \dotfill & $   -56.84 $ & $     2.23 $ & $     5.54 $ & $     4.57 $ & $    0.0046 $   \\
$ 1319.12199 $ \dotfill & $   -57.72 $ & $     2.40 $ & $    -5.57 $ & $     5.53 $ & $    0.0048 $   \\
$ 1321.12276 $ \dotfill & $   -20.39 $ & $     1.37 $ & $     1.22 $ & $     5.01 $ & $    0.0047 $   \\
$ 1351.97316 $ \dotfill & $   -26.02 $ & $     1.50 $ & \nodata      & \nodata      & \nodata          \\

%% file: phfu_tab_short.tex
$ 54449.55537 $ & $  -0.00357 $ & $   0.00095 $ & $   9.35242 $ & $ z$\\
$ 54449.55568 $ & $   0.00062 $ & $   0.00094 $ & $   9.35614 $ & $ z$\\
$ 54449.55603 $ & $  -0.00175 $ & $   0.00095 $ & $   9.35430 $ & $ z$\\
$ 54449.55633 $ & $   0.00203 $ & $   0.00095 $ & $   9.35788 $ & $ z$\\
$ 54449.55668 $ & $  -0.00602 $ & $   0.00093 $ & $   9.35026 $ & $ z$\\
$ 54449.55700 $ & $   0.00173 $ & $   0.00095 $ & $   9.35703 $ & $ z$\\
$ 54449.55732 $ & $  -0.00628 $ & $   0.00093 $ & $   9.34882 $ & $ z$\\
$ 54449.55768 $ & $  -0.00183 $ & $   0.00094 $ & $   9.35300 $ & $ z$\\
$ 54449.55799 $ & $  -0.00084 $ & $   0.00094 $ & $   9.35409 $ & $ z$\\
$ 54449.55835 $ & $   0.00139 $ & $   0.00093 $ & $   9.35711 $ & $ z$\\

%% file: biblio.tex
%%%% %T\n %%%% %u\n \\bibitem[%\3m%(y)]{%\1h:%\Y} %\8l~%\Y,%\j,%\V,%\p\n

%% \%% %3.3a\n\%% %t\n\%% %R\n\%%%u\n\\bibitem[%\4m%(y)]{%\1H%\Y} 
%\8l~%\Y,%\j,%\V,%\p\n